\documentclass[a4paper,11pt]{article}
\pdfoutput=1 

\usepackage{jheppub} 
\usepackage[T1]{fontenc}
\usepackage{lmodern}
\usepackage{bbold}
\usepackage{amsmath}
\usepackage{amssymb}
\usepackage{graphicx}
\usepackage{xspace}
\usepackage{slashed}
\usepackage{subcaption}
\newcommand{\nn}{\nonumber} 
\newcommand{\bn}{{\bar n}}

\newcommand*\diff{\mathop{}\!\mathrm{d}}

\newcommand{\be}{\begin{equation}}
\newcommand{\ee}{\end{equation}}
\newcommand{\bea}{\begin{eqnarray}}
\newcommand{\eea}{\end{eqnarray}}

\newcommand{\ma}{\mathrm}
\newcommand{\ml}{\mathcal}
\newcommand{\bs}{\boldsymbol}

\newcommand{\cL}{\mathcal{L}}

\newcommand{\cB}{\mathcal{B}}
\newcommand{\cP}{\mathcal{P}}

\allowdisplaybreaks[1]

\DeclareMathOperator{\Tr}{Tr}

\title{\boldmath Transverse Momentum Broadening of a Jet in Quark-Gluon Plasma: An Open Quantum System EFT}

\author{Varun Vaidya}
\author{and Xiaojun Yao}

\affiliation{Center for Theoretical Physics, Massachusetts Institute of Technology \\ Cambridge, MA 02139, USA}

\emailAdd{vvaidya@mit.edu}
\emailAdd{xjyao@mit.edu}

\abstract{We utilize the technology of open quantum systems in conjunction with the recently developed effective field theory for forward scattering to address the question of massless jet propagation through a weakly-coupled quark-gluon plasma in thermal equilibrium. We discuss various possible hierarchies of scales that may appear in this problem, by comparing thermal scales of the plasma with relevant scales in the effective field theory. Starting from the Lindblad equation, we derive and solve a master equation for the transverse momentum distribution of a massless quark jet, at leading orders both in the strong coupling and in the power counting of the effective field theory. Markovian approximation is justified in the weak coupling limit. Using the solution to the master equation, we study the transverse momentum broadening of a jet as a function of the plasma temperature and the time of propagation. We discuss the physical origin of infrared sensitivity that arises in the solution and a way to handle it in the effective field theory formulation. We suspect that the final measurement constraint can only cut-off leading infrared singularities and the solution to the Markovian master equation resums a logarithmic series. This work is a stepping stone towards understanding jet quenching and jet substructure observables on both light and heavy quark jets as probes of the quark-gluon plasma.}

\preprint{MIT-CTP/5187}

\begin{document} 
\maketitle
\flushbottom

\section{Introduction}
Jets are sprays of collimated particles produced in high energy collisions of hadrons and/or electrons. Their formation starts with a highly virtual parton generated from an initial hard scattering, followed by subsequent parton cascade and fragmentation. Jet production can be studied via perturbative QCD due to the large scales involved, for example, the virtuality of the initial parton produced. Therefore the calculation of the initial production of jets can be well-controlled theoretically, which makes jets powerful tools to probe the properties of the quark-gluon plasma (QGP) in heavy ion collisions.

Jet production is modified in heavy ion collisions, compared with that in proton-proton collisions, due to the jet-medium interaction. Jet quenching, a phenomenon of suppression of particles with high transverse momenta, has been studied intensively theoretically long before \cite{Gyulassy:1993hr,Wang:1994fx,Baier:1994bd,Baier:1996kr,Baier:1996sk,Zakharov:1996fv,Zakharov:1997uu,Gyulassy:1999zd,Gyulassy:2000er,Wiedemann:2000za,Guo:2000nz,Wang:2001ifa,Arnold:2002ja,Arnold:2002zm,Salgado:2003gb,Armesto:2003jh,Majumder:2006wi,Majumder:2007zh,Neufeld:2008fi,Neufeld:2009ep} and recently observed in experiments at both Relativistic Heavy Ion Collider (RHIC) \cite{Arsene:2004fa,Back:2004je,Adams:2005dq,Adcox:2004mh} and Large Hadron Collider (LHC) \cite{Aad:2010bu,Aamodt:2010jd,Chatrchyan:2011sx}. The suppression mechanism is mainly the energy loss when jets traverse the hot medium. Both collisional and medium-induced radiative energy loss contribute, but the latter dominates at high energy. The key to understand jet quenching and jet substructure modifications in heavy ion collisions is to understand how the jet interacts with the expanding medium. There has been tremendous theoretical effort to study the jet energy loss mechanism (see Refs.~\cite{Mehtar-Tani:2013pia,Blaizot:2015lma,Qin:2015srf,Cao:2020wlm} for recent reviews). But this is not a simple problem because it involves multiple scales such as the jet energy, the transverse momentum with respect to the jet axis and thermal scales of the QGP. Furthermore, in current heavy ion collision experiments, the temperature achieved fits roughly the range $150 - 500$ MeV, and may not always be a perturbative scale. Thus, a fully weak coupling calculation may not be valid. A hybrid model has been developed to address this problem \cite{Casalderrey-Solana:2014bpa,Casalderrey-Solana:2015vaa,Hulcher:2017cpt,Casalderrey-Solana:2018wrw,Casalderrey-Solana:2019ubu}, in which the initial jet production and vacuum-like parton shower are calculated perturbatively, while the subsequent jet energy loss in the medium is calculated by mapping the field theory computation in the strong coupling limit to a weak coupling computation in the classical gravity theory \cite{Liu:2006ug,Argyres:2006yz,CasalderreySolana:2007qw,Hatta:2008tx,Chesler:2008uy,DEramo:2010wup}, i.e., by using the AdS/CFT correspondence \cite{Maldacena:1997re}. 

From the perspective of field theory, a powerful tool to deal with multi-scale problems is effective field theory (EFT). The EFT that is particularly useful for jet studies is Soft-Collinear Effective Theory (SCET). There are also formulations of SCET (known as SCET$_G$) treating the Glauber gluon, which is a type of mode appearing in forward scattering, as a background field induced by the medium interacting with an energetic jet. By making use of the collinear sector of the corresponding  EFT, this formalism has been used to address the question of jet quenching in the medium \cite{Ovanesyan:2011kn,
Chien:2015hda,
Ovanesyan:2011xy,
Chien:2015vja,
Kang:2014xsa}. In the same spirit, a new EFT for forward scattering has been developed recently \cite{Rothstein:2016bsq} which also uses the Glauber mode to write down contact operators between the soft and collinear momentum degrees of freedom. The partons from the thermal QGP are generally soft, when compared with an energetic jet, which can be described by a collinear mode.

Another theoretical challenge in understanding the jet-medium interaction is the quantum interference effect. For example, in the process of the medium-induced single radiation from a high energy parton, multiple transverse momentum kicks from the medium can suppress the radiation spectrum by destructive interference, a phenomenon known as the Landau-Pomeranchuk-Migdal (LPM) effect. Progress in understanding the LPM effect in the single radiation has been achieved in recent years \cite{CaronHuot:2010bp,Ke:2018jem,Mehtar-Tani:2019ygg}. Extension to studying the LPM effect in multiple splittings has been explored in simplified cases \cite{Arnold:2015qya,Arnold:2016kek}. Furthermore, when two collimated partons are close to each other spatially, the medium may not be able to resolve them completely. So they may lose energy coherently as a single parton. This interference effect caused by the finite resolution power of the QGP is also important and can change jet substructure observables dramatically \cite{Casalderrey-Solana:2019ubu}.

To take into account the interference effect systematically, one can keep track of the time evolution of the system's density matrix. This can most easily be done by using the open quantum systems formalism (for introductory books, see \cite{Breuer:2002pc,OQS}). For jets inside a QGP, if we only focus on jet observables, the jet can be treated as an open quantum system interacting with a QGP bath. The application of the open quantum system formalism in heavy ion collisions has been thriving in the study of color screening and regeneration of quarkonium \cite{Young:2010jq,Borghini:2011ms,Akamatsu:2011se,Akamatsu:2014qsa,Blaizot:2015hya,Katz:2015qja,Kajimoto:2017rel,DeBoni:2017ocl,Blaizot:2017ypk,Blaizot:2018oev,Akamatsu:2018xim,Miura:2019ssi}. There, the heavy quark-antiquark pair in the color singlet interacts with the medium destructively when they are close. Great progress in the understanding of quarkonium in-medium dynamics has been achieved by combining potential nonrelativistic QCD (pNRQCD \cite{Brambilla:1999xf,Brambilla:2004jw,Fleming:2005pd}, an EFT of QCD) and the open quantum system formalism \cite{Brambilla:2016wgg,Brambilla:2017zei,Brambilla:2019tpt,Yao:2018nmy}. For example, a semiclassical Boltzmann transport equation of quarkonium in the medium has been derived, under assumptions that are closely related with a hierarchy of scales \cite{Yao:2018nmy,Yao:2019jir,Yao:2020kqy}.

We would like to combine the forward scattering EFT recently developed within the formalism of SCET with the open quantum system formalism and explore its physical implications on the jet-medium interaction. As a first step, we will study in this paper, the transverse momentum broadening of a high energy parton moving through the medium. For simplicity, we will assume the plasma temperature is high enough so the weak coupling calculation is valid. We will leave the inclusions of nonperturbative effects and radiation into the calculation to future work.

The use of an EFT formalism allows us to write a simple and hence easily calculable description of the system. On the other hand, as alluded to earlier, we hope that the use of an open quantum system approach will allow us to easily keep track of the quantum interference effects and construct a new class of calculable observables. 
The long term goal here is to develop a theoretically robust formalism for computing jet substructure observables for both light parton and heavy quark jets. For example, the bottom quark jets have been identified as an effective probe of the QGP medium and will be experimentally studied at LHC, as well as by the sPHENIX collaboration at RHIC. There has been recent work on computing jet substructure observable for heavy quark jets in the context of proton-proton collisions \cite{Lee:2019lge,Makris:2018npl}. The objective would then be to compute the same observables in heavy ion collisions and study modifications caused by the medium.

This paper is organized as follows.
In Section~\ref{sec:Hierarchy} we discuss the importance of the forward scattering regime in jet-medium interactions. Based on this, we then introduce the physical system that we wish to study and the relevant physical scales that play an important role in its description.
The next Section~\ref{sec:SCET} reviews the basics of SCET and Glauber EFTs as tools to model jet propagation in the QGP. We then introduce the concept of open quantum systems and derive a master equation for the jet density matrix from the Lindblad equation in Section~\ref{sec:OpenQSystem}. The master equation is solved analytically and the transverse momentum distribution of a jet is studied in Section~\ref{sec:Transverse} along with a comparison with previous results in literature. We also discuss possible infrared (IR) divergences that show up in the Markovian limit. Finally we conclude and discuss future directions in Section~\ref{sec:conclusions}.

\section{Relevant Hierarchy of Scales}
\label{sec:Hierarchy}In this section we examine the dominant interaction of a jet with a QGP medium and discuss the possible hierachy of scales that can appear as a function of the jet energy and QGP temperature.

\subsection{Dominance of Forward Scattering Regime}
\label{sec:Forward scattering}
We can gain intuition about the dominant interaction of a parton traversing a QGP medium by examining a $2\to2$ scattering.  
Consider the simple example of $2\rightarrow2$ scattering $e^-\mu^-\rightarrow e^-\mu^-$. The lowest order Feynman diagram is just a $t$ channel photon exchange. The differential cross section for this process has the form \cite{Peskin}
\bea
\frac{d\sigma}{d\Omega}=\frac{\alpha_\ma{EM}^2}{2E^2_\ma{cm}(1-\cos \theta)^2}\left(4+(1+\cos \theta)^2\right) \,,
\eea
where $E_\ma{cm}$ is the center of mass energy and $\theta$ is the angle between the final and initial state electron/muon. $\alpha_\ma{EM}$ is the electromagnetic coupling.
This cross section has a singular behavior as $\theta \rightarrow 0$
\bea
\frac{d\sigma}{d\Omega}\propto \frac{1}{\theta^4} \,.
\eea
This singularity over the phase space is not integrable and is in fact a physical singularity that arises due to the infinite range of the Coulomb potential. In real experiments, however, this singularity gets cut off by some IR scale such as a dynamically induced photon mass or a finite interaction region introduced by localized beam wave packets at a finite impact parameter. It is worth noting that most of the contribution to the cross section comes from this small angle region of phase space. In the case of scattering inside the QGP, the corresponding $2\rightarrow2$ scattering would be forward scattering of quarks/gluons mediated by a gluon. For a QGP in thermal equilibrium with a temperature $T$, the interactions in the medium induce an effective gluon mass (Debye mass) $m_D$ which acts as an IR cut-off scale. As we will do in this paper, we may also impose some cuts on the final state measurements which will act to regulate some of the IR divergences.

We wish to develop an EFT description in this region of forward scattering by expanding in the small scattering angle $\theta$, which will also be the power counting parameter of our EFT. How exactly this parameter is related to the physical scales of the system will depend on the measurement that we impose on the final state. Roughly speaking, for a high energy parton (quark or gluon) with an initial energy $Q$, the angular parameter $\theta \sim Q_{\perp}/Q$ measures the transverse momentum of the final state with respect to its initial direction (considered as the longitudinal direction). We will discuss this in detail in the following subsection.


\subsection{Jets in Quark-Gluon Plasma}
The QGP at vanishing chemical potential\footnote{We leave the case of a  non-zero baryon density to future studies.} is characterized by its temperature $T$ in thermal equilibrium. In this paper we will mostly be concerned with light quarks so all the partons are considered massless at the level of the Lagrangian. The finite temperature interactions between the partons induce a dynamical gluon mass $m_D$, which is of the order of $gT$ in perturbation. Here $g$ is the strong coupling at the scale $T$. The Debye mass provides a screening effect and effectively shortens the strong interaction range to be of the order of $1/m_D$. The perturbative description of the QGP works well when the temperature $T$ is far above the confinement scale $\Lambda_\ma{QCD}$, which is a dynamically generated scale of QCD.

The system we want to study is a highly energetic jet traversing a region of the QGP. The energy of the jet, $Q$, will be the hard scale in our process and is assumed to be much larger than all the other scales in the problem. In this paper, we will focus on the leading interaction of the jet with the QGP medium and leave the vacuum as well as medium-induced splitting to future studies. So effectively our jet is described by a leading parton. The aim is to calculate the final transverse momentum broadening of the jet when it comes out of the QGP. For our purpose in this paper, we are going to impose a measurement constraint on the final transverse momentum $Q_{\perp}$ of the jet, by concentrating on the forward scattering region, i.e., $\lambda \sim \theta \ll 1$. Here $\lambda$ is the power counting parameter for our EFT description. If the interaction of the jet with the medium is a single coherent scattering, then we can write $\lambda \sim \theta \sim Q_{\perp}/Q$. In this paper, however, we allow the jet to have a series of mutually incoherent interactions with the medium.  In any of the intermediate incoherent interactions, a smaller transverse momentum $p_{\perp} \lesssim Q_\perp$ can be exchanged between the jet and the medium so along as the net value adds up to $Q_{\perp}$. This means that our power counting parameter satisfies $ \lambda \lesssim Q_{\perp}/Q$, with the smallest transverse momentum being effectively cut-off by $m_D$. We will discuss this in detail in Section \ref{sec:Transverse}. Since the transverse momentum exchanged is always small compared to the jet energy, the jet (leading parton) can be treated as a collinear particle throughout its evolution. Medium partons will be treated as soft modes that carry energy and momentum of the order of the QGP temperature $T$. We are always going to work in a regime where $T \ll Q$.

We can use the final measurement on $Q_\perp$ to probe the physics at different scales (while maintaining $Q_\perp/Q \ll 1$, otherwise our EFT framework described below does not apply). 
The natural question then is how the IR scale $Q_{\perp}$ compares to all the other scales such as $T$, $m_D$ and $\Lambda_\ma{QCD}$. Several possible hierarchies are possible here and we now discuss each of them.

\begin{itemize}
    \item{High temperature}
    
    The simplest case is when the temperature $T$ is high enough that both $T$ and $m_D$ are perturbative scales. In other words, we have $Q \gg T \gg m_D \gg \Lambda_\ma{QCD}$. 
     We can also have the possibility of a somewhat lower temperature such that $Q \gg T \sim m_D \gg \Lambda_\ma{QCD}$, where the Debye mass scale is still perturbative. Under these hierarchies, we can do a perturbative computation of the final observable $Q_\perp$ to probe all the scales greater than $\Lambda_\ma{QCD}$ and much smaller than Q. We will primarily focus on this regime in this paper. 
     
    \item{Intermediate temperature}
    
    In this case, we still have $T$ high enough to be a perturbative scale. However, the scale $m_D \sim \Lambda_\ma{QCD}$ is now nonperturbative, i.e., $Q\gg T \gg m_D\sim \Lambda_\ma{QCD}$. But we can still set up a perturbative EFT for $Q_{\perp}\sim T$. In fact the EFT that we construct in the high $T$ regime in the first case will be valid for the $Q_{\perp}\sim T$ case here as well. On the other hand, if we want to probe the physics at lower scales, $Q_{\perp}\sim m_D$, then we have to take into account nonperturbative effects.
    
    \item{Low temperature}
    
    Finally we have the low temperature regime in which case all our scales $T\sim m_D \sim \Lambda_\ma{QCD}$ are nonperturbative. In this case, we can still develop a perturbative EFT description for the case $Q \gg Q_{\perp} \gg T$, and then appropriately make a transition to the nonperturbative regime of $Q_{\perp}$. 
\end{itemize}

As a side remark, we want to emphasize that jet observables with the same jet radius $R$ can be very different at the RHIC and LHC energies. Since at LHC the collision energy is much higher, more energetic jets can be produced, whose energies $Q$ can be much larger than those at RHIC with the same jet radius. Then the transverse momenta $\sim QR$ (with respect to the jet axis) of jets at RHIC and LHC can be very different, even though these jets are defined with the same jet radius $R$, and probably probe the physics at different scales. For example, it is likely that the transverse momentum at the LHC energy is in the perturbative regime while that at the RHIC energy sits in the nonperturbative regime. EFT approaches can help us to better understand the difference quantitatively.

Since we will concentrate on the region of forward scattering, we can use EFT tools already available in the literature to describe our system. One such a formalism that has been extensively used in collider physics is SCET. We now review the basics of SCET and discuss how it can be used in our problem.


\section{Soft-Collinear Effective Theory(SCET) and Forward Scattering}
\label{sec:SCET}

\subsection{Review of Soft-Collinear Effective Theory}
SCET is a theory of both soft and collinear particles. Collinear particles have a large momentum along a particular light-like direction, while soft particles have a small momentum, and no preferred direction.  For each relevant light-like direction, we define two reference vectors $n^\mu$ and $\bn^\mu$ such that $n^2 = \bn^2 = 0$ and $n\cdot \bn = 2$.  The typical choice of $n^{\mu}=(1,0,0,1)$ and $\bar n^{\mu}=(1,0,0,-1)$ will be used below. The freedom in the choice of $n$, as in the case of the label velocity in Heavy Quark Effective Theory, is represented in the EFT by a reparametrization invariance ~\cite{Manohar:2002fd,Chay:2002vy}.   Any four-momentum $p$ can be decomposed with respect to $n^\mu$ as
\begin{equation} 
\label{eq:lightcone_dec}
p^\mu = \bn\cdot p\,\frac{n^\mu}{2} + n\cdot p\,\frac{\bn^\mu}{2} + p^\mu_{\perp}\
\,.
\end{equation}

The SCET is defined by a systematic expansion in terms of a formal power counting parameter $\lambda \ll 1$, which is determined by the measurements or kinematic restrictions imposed on the QCD radiation. The momenta for different modes in the SCET scale as
\bea
\text{Collinear} &:& ~\big(\bn \!\cdot\! p, n\!\cdot\! p, p_{\perp}\big) \sim \bn\!\cdot\! p \,\big(1,\lambda^2,\lambda\big)\,, \nn \\[2pt]
\text{Soft} &:& ~\big(\bn \!\cdot\! p, n\!\cdot\! p, p_{\perp}\big) \sim \bn\!\cdot\! p \,\big(\lambda,\lambda,\lambda\big)\,,  \\[2pt]
\text{Ultrasoft} &:& ~\big(\bn \!\cdot\! p, n\!\cdot\! p, p_{\perp}\big) \sim \bn\!\cdot\! p \,\big(\lambda^2,\lambda^2,\lambda^2\big)\,.\nn
\eea
 A theory with only collinear and ultrasoft modes is typically referred to as SCET I, while that with only collinear and soft modes is referred to as SCET II \cite{Bauer:2002aj}\footnote{In the presence of Glauber modes, soft modes are always required for the renormalization group consistency of the Glauber potentials \cite{Rothstein:2016bsq}. Whether or not ultrasoft modes are required depends on the physical observable in question.}. In this paper, we will only be concerned with  SCET II along with the Glauber mode.

In order to expand the fields in the full theory QCD around a particular direction, the momenta are decomposed into the label $\tilde{p}^\mu$ and residual $k^\mu$ components
\begin{equation} \label{eq:label_dec}
p^\mu = \tilde p^\mu + k^\mu = \bn \cdot \tilde p\, \frac{n^\mu}{2} + \tilde p_{\perp}^\mu + k^\mu\,.
\,\end{equation}
Then for a collinear particle, $\bn \cdot\tilde p \sim Q$ and $\tilde p_{\perp} \sim \lambda Q$, where $Q$ is a typical scale of the hard interaction, while $k^\mu \sim \lambda^2 Q$ describes small fluctuations around the label momentum. Field modes with momenta of definite scaling in the SCET are obtained by performing a multipole expansion of the fields in the full theory QCD. SCET involves independent gauge bosons and fermions for each collinear direction $A_{n,\tilde p}(x)$, $\xi_{n, \tilde p}(x)$, which are labeled by their collinear direction $n$ and their large label momentum $\tilde p$, as well as (ultra)soft gauge boson fields $A_{(u)s}(x)$. Independent gauge symmetries are enforced for each set of fields, which have support for the corresponding momentum carried by that field \cite{Bauer:2003mga}. Overlap between different regions is removed by the zero-bin subtraction procedure \cite{Manohar:2006nz}. This ensures no double counting of momentum regions. 

The leading power SCET II Lagrangian that we shall be concerned with in the following, takes the form
\begin{equation} 
\label{eq:SCETLagExpand}
\cL_{\text{SCET}}=  \cL_h^{(0)}+\cL_{c}^{(0)}+ \cL_{s}^{(0)}+ {\cal L}_G^{(0)} \,.
\end{equation}
Here $ \cL_h^{(0)}$ contains the hard scattering operators and is determined by an explicit matching calculation. The Lagrangian $\cL_c^{(0)}$, $\cL_s^{(0)}$ describe the universal leading power dynamics of the collinear and soft modes. Finally, ${\cal L}_G^{(0)} $ is the leading power Glauber Lagrangian \cite{Rothstein:2016bsq}, which describes the leading power coupling between the soft and collinear degrees of freedom through potential operators. We will discuss the Glauber interaction in more detail in the next section.

Hard scattering operators involving collinear fields are constructed out of products of Wilson line dressed fields that are invariant under collinear gauge transformations~\cite{Bauer:2000yr,Bauer:2001ct}.  For example, the gauge invariant gauge boson operator is given by
\begin{align} \label{eq:chiB}
\cB_{n}^\mu(x)
= \frac{1}{g}\Bigl  [W_{n}^\dagger(x)\,i  D_{{n}\perp}^\mu W_{n}(x)\Bigr]
 \,, 
\end{align}
where $D_{n\perp}$ is the collinear gauge covariant derivative, and $W_n(x)$ is a collinear Wilson line 
\bea
W_n(x) =\left[ \, \sum\limits_{\text{perms}} \exp \left(  -\frac{g}{\bar{n}\cdot \cP} \bar n \cdot A_n(x)  \right) \right]\,,
\eea
where $\cP^\mu$ is an operator that returns the label momentum of the fields on its right. The collinear Wilson line, $W_{n}(x)$, is localized with respect to the residual position $x$ so that $\cB_{{n}}^\mu(x)$ can be treated as local gauge boson fields from the perspective of the ultrasoft degrees of freedom.  For the leading power calculation presented here, ultrasoft and soft fields will not appear explicitly in our hard scattering operators, other than through the Wilson lines in the field redefinition
\begin{align} \label{eq:BPSfieldredefinition}
\cB^{A\mu}_{n\perp}\to Y_n^{AB} \cB^{B\mu}_{n\perp}\,,
\end{align}
which is performed on each collinear sector. For a general representation, $r$, the ultrasoft Wilson line is defined by\footnote{Here we give the explicit result for an incoming Wilson line. Depending on whether particles are incoming or outgoing, different Wilson lines must be used. When done correctly, the BPS field redefinition accounts for the full physical path of the particles \cite{Chay:2004zn,Arnesen:2005nk}.}
\begin{align}
Y^{(r)}_n(x)=\bold{P} \exp \left [ ig \int\limits_{-\infty}^0 \diff s\, n\cdot A^B_{us}(x+sn)  T_{(r)}^{B}\right]\,,
\label{eq:YWilsonLine}
\end{align}
where $\bold P$ denotes path ordering.  This so-called BPS field redefinition has the effect of decoupling ultrasoft and collinear degrees of freedom at leading power \cite{Bauer:2002nz}, and it accounts for the full physical path of ultrasoft Wilson lines~\cite{Chay:2004zn,Arnesen:2005nk}. In the following, we will also need soft Wilson lines,
\begin{align}
S^{(r)}_n(x)=\bold{P} \exp \left [ ig \int\limits_{-\infty}^0 \diff s\, n\cdot A^B_{s}(x+sn)  T_{(r)}^{B}\right]\,.
\end{align}


\subsection{Glauber Mode for Forward Scattering}
The process of near forward scattering is referred to as the Glauber exchange which involves the exchange of an off-shell gluon. The transverse momentum of this gluon (with respect to the forward direction) is parametrically larger than its longitudinal components so that $|k_{\perp}|^2 \gg \bar n \cdot k n \cdot k$, which is different from the regime of a Coulomb exchange which satisfies $|{\bs k}|^2 \gg (k^0)^2$. As a result, the Glauber mode is not a propagating mode and acts instantaneously along the light-cone time.

A systematic study of the Glauber mode was carried out in \cite{Rothstein:2016bsq} within the formalism of SCET. Depending on the process that we are interested in, the asymptotic (propagating) states that we deal with can be classified as collinear and soft modes as described in the previous section. Scattering processes at colliders can usually be factorized in terms of these modes that separate the momentum fluctuations at different scales. The Glauber is an off-shell mode that mediates between either two collinear, two soft or one collinear and one soft modes, thus violating factorization. These factorization violating interactions can be captured via effective operators in the Glauber Lagrangian $\mathcal{L}^{(0)}_G$. At leading power, these operators have been derived in \cite{Rothstein:2016bsq}.

Our interest lies in utilizing this EFT formalism to study the near forward scattering of a jet inside a QGP in thermal equilibrium. For the $2\to2$ scattering process, the region of near forward scattering dominates the total cross section and various approximations can be made to simplify the result at leading power in the expansion parameter. In the current case the expansion parameter is the small scattering angle $\theta$.


We can now discuss the effective interaction between the jet leading parton and the medium in the forward scattering region. The thermal QGP is mainly composed of soft particles whose energies and momenta are on the order of $T$. Their momenta $p_s$ scale uniformly in our expansion parameter $\lambda \sim \theta \ll 1$, so we can write
\bea
p_s \sim Q( \lambda, \lambda, \lambda)\,,
\eea
where $Q$ is the hard scale in our process, i.e., the jet energy.
The jet is composed of highly energetic collinear particles that are moving along the light-like direction $n$. The scaling of their momenta, in the light-cone coordinate, can be written as 
\bea 
p_c \sim Q(1, \lambda^2, \lambda) \,.
\eea
As the collinear particles move through the medium, they interact with the soft medium particles mainly via forward scattering where both the collinear and soft particles maintain their momentum scaling after the scattering. This interaction is therefore mediated by the Glauber mode with the scaling  
\bea
p_G \sim Q(\lambda, \lambda^2, \lambda) \,.
\eea
Adding or subtracting a momentum of the Glauber scaling from the collinear or soft mode does not alter their momentum scaling.

The Glauber can be integrated out of the Lagrangian, which leads to effective operators coupling the collinear and soft degrees of freedom.
The effective gauge invariant operators for quark-quark ($qq$), quark-gluon ($qg$ or $gq$) and gluon-gluon ($gg$) interactions have been worked out in the Feynman gauge in Ref.~\cite{Rothstein:2016bsq}
\bea
\label{EFTOp}
\mathcal{O}_{ns}^{qq}&=&\mathcal{O}_n^{qB}\frac{1}{\mathcal{P}_{\perp}^2}\mathcal{O}_s^{q_nB} \,, \nn\\
 \mathcal{O}_{ns}^{qg}&=&\mathcal{O}_n^{qB}\frac{1}{\mathcal{P}_{\perp}^2}\mathcal{O}_s^{g_nB} \,, \nn\\
\mathcal{O}_{ns}^{gq}&=&\mathcal{O}_n^{gB}\frac{1}{\mathcal{P}_{\perp}^2}\mathcal{O}_s^{q_nB} \,,\nn\\
\mathcal{O}_{ns}^{gg}&=&\mathcal{O}_n^{gB}\frac{1}{\mathcal{P}_{\perp}^2}\mathcal{O}_s^{g_nB}\,,
\eea
where $B$ is the color index and the subscripts $n$ and $s$ denote the collinear and soft operators respectively. The soft operators $\mathcal{O}_s$ are constructed from the gauge invariant soft quark and gluon building blocks that are built out of the soft fields dressed with soft Wilson lines:
\bea
\mathcal{O}_s^{q_nB} &=& 8\pi \alpha_s\left(\bar \psi^n_s T^B \frac{\slashed{n}}{2}\psi_s^n\right) \,,\nn\\
\psi_s^n &=& S_n^{\dagger}\psi_s \,, \nn\\ 
\mathcal{O}^{g_nB}_s &=& 8\pi \alpha_s\left(\frac{i}{2}f^{BCD}\mathcal{B}_{s\perp }^{nC}\frac{n}{2}\cdot (\mathcal{P}+\mathcal{P}^{\dagger}) \mathcal{B}_{s\perp}^{nD} \right) \,, \nn\\
\label{eqn:softO}
\mathcal{B}_{s\perp}^{n\mu} &=& \mathcal{B}_{s\perp}^{nB\mu} T^B =  \frac{1}{g}\big(S_n^{\dagger}iD_{s\perp}^{\mu} S_n\big) \,,
\eea
where the soft Wilson lines ensure that the operators are invariant under soft gauge transformations.

The collinear operators are built out of the collinear building blocks. In this paper, we will only work with collinear quarks which are constructed from bare collinear quark fields dressed with collinear Wilson lines:
\bea
\mathcal{O}_n^{qB} &=& \bar{\chi}_n T^B \frac{\slashed{\bar n}}{2} \chi_n \\
\chi_n &=& W_n^\dagger \xi_n = W_n^\dagger \frac{\slashed{n}\slashed{\bar{n}}}{4}\psi \,,
\eea
where $\psi$ is the standard four-component Dirac spinor.
Since the soft momentum puts the collinear particle off-shell and off-shell modes have been integrated our in the construction of the EFT, the collinear fields do not transform under the soft gauge transformations.

The effective Lagrangian density for the Glauber exchange then looks like 
\bea
\mathcal{L}_G = e^{-ix\cdot \mathcal{P}}\sum \mathcal{O}_i  \,,
\eea
in which the operators $\ml{O}_i$ are listed in Eq.~(\ref{EFTOp}).

One point to note here is that the collinear and soft operators $\mathcal{O}^n$ and $\mathcal{O}^s$ are separately gauge invariant. However, a simple calculation shows that a change in the gauge choice for the Glauber propagator in the construction would lead to a different form of the gauge invariant operators. The derivation of Eq.~(\ref{EFTOp}) in Ref.~\cite{Rothstein:2016bsq} chooses the Feynman gauge. In general, the operators will be of the form
\bea
\mathcal{O}_{n \mu} \big[\Delta^{\mu \nu} \big] \mathcal{O}_{s\nu} \,,
\eea 
where $\Delta^{\mu \nu}$ would be the Glauber gluon propagator in the chosen gauge. The operators $\mathcal{O}_{n \mu}, \mathcal{O}_{s\mu}$ would still be separately gauge invariant \footnote{The operator $O_{n \mu}$ is invariant under collinear gauge transformations while $O_{s\nu}$ is invariant under Soft gauge transformation.}  but the form of these operators would change according to the Glauber gauge choice. The final result for any scattering amplitude would remain gauge independent. Thus it suffices to work with any specific gauge. From now on we will choose to work in the Feynman gauge since all the effective operators have been constructed.


\section{Lindblad Equation for Open Quantum System}
\label{sec:OpenQSystem}
In this section we review the basic concepts of the open quantum system formalism, which can be used to describe the quantum dynamical evolution of an open subsystem in contact with an environment. Later in this section, we will apply this formalism to the case of a jet (subsystem) interacting with the thermal QGP (environment) via the effective operators described in the previous section.

We begin with the microscopic derivation of the Lindblad equation which closely follows the discussion in Ref.~\cite{OQS}. We assume that the Hamiltonian of the total system (subsystem and environment) is given by
\be
H = H_S +H_E + H_I\,,
\ee
where $H_S$ is the subsystem Hamiltonian, $H_E$ is the environment Hamiltonian, and $H_I$ contains the interactions between the subsystem and the environment. The interaction Hamiltonian is assumed to be factorized as follows: $H_I = \sum_{\alpha} {O}^{(S)}_{\alpha} \otimes {O}^{(E)}_{\alpha}$ where ${O}^{(S)}_{\alpha}$ and ${O}^{(E)}_{\alpha}$ denote the subsystem and environment operators respectively. Here $\alpha$ denotes all relevant quantum numbers. (For local quantum field theory, the factorized form is generally true and $\alpha$ includes the spatial coordinates for which the summation means an integration.) In our case, $\alpha$ would be the particle type ($q$ or $g$), the color and spatial coordinates of the effective operators. We can assume $\langle O^{(E)}_{\alpha}\rangle \equiv \Tr_E(O^{(E)}_{\alpha}\rho_E) = 0$ because we can redefine $O^{(E)}_{\alpha}$ and $H_S$ by $O^{(E)}_{\alpha} - \langle O^{(E)}_{\alpha}\rangle$ and $H_S + \sum_{\alpha} O^{(S)}_{\alpha} \langle O^{(E)}_{\alpha}\rangle $ respectively. Here $\rho_E$ is the density matrix of the environment. Each part of the Hamiltonian is assumed to be Hermitian.

The von Neumann equation for the time evolution of the total density matrix in the interaction picture is given by
\be
\frac{\diff \rho^{(\ma{int})}(t)}{\diff t} = -i [H^{(\ma{int})}_I(t), \rho^{(\ma{int})}(t)] \,,
\ee 
where
\bea
 \rho^{(\ma{int})}(t) &=& e^{iH_S t}e^{iH_E t}\rho(t)e^{-iH_E t}e^{-iH_S t} \,, \\
 H^{(\ma{int})}_I(t) &=& e^{iH_St}e^{iH_Et} H_I(t) e^{-iH_E t}e^{-iH_S t} \,.
\eea
The density matrix and Hamiltonians without any superscript are in the Schr\"odinger picture. In the above formula, we have used the fact $[H_S, H_E]=0$.
We will omit the superscript ``(int)'' in the following discussion. The symbolic solution is given by
\be
\rho(t) = U(t) \rho(0) U^{\dagger}(t)\,,
\ee
where the evolution operator is
\be
U(t) = \ml{T} e^{-i \int_0^t H_I(t') \diff t'}\,,
\ee
and $\ml{T}$ is the time-ordering operator.

We will assume the subsystem and the environment are weakly interacting. We further assume the initial total density matrix factorizes
\be
\rho(0) = \rho_S(0) \otimes \rho_E \,,
\ee
which is generally true for weakly-coupled systems (factorization breaking terms come at higher orders in the coupling). The environment density matrix is assumed to be in thermal equilibrium:
\be
\rho_E = \frac{e^{-\beta H_E}}{\Tr_E e^{-\beta H_E}} \,,
\ee
where $T = 1/\beta$ is the temperature of the thermal environment. If we expand the interaction to second order in perturbation and take the partial trace over the environment degrees of freedom, we obtain the Lindblad equation:
\bea
\label{eqn:lindblad}
\rho_S(t) = \rho_S(0) & -&i\sum_{a,b} \sigma_{ab}(t) [L_{ab}, \rho_S(0)] \\ \nn
& + & \sum_{a,b,c,d} \gamma_{ab,cd} (t) \Big( L_{ab}\rho_S(0)L^{\dagger}_{cd} - \frac{1}{2}\{ L^{\dagger}_{cd}L_{ab}, \rho_S(0)\}  \Big)
 + \ml{O}\big( (H_I)^3 \big)\,.
\eea
Each term in the Lindblad equation is defined as
\bea
L_{ab} &\equiv& |a\rangle \langle b| \\
\sigma_{ab}(t) &\equiv& \frac{-i}{2} \sum_{\alpha, \beta} \int_0^t \diff t_1 \int_0^t \diff t_2 C_{\alpha\beta}(t_1, t_2) \text{sgn}(t_1-t_2) \langle a | O^{(S)}_{\alpha}(t_1) O^{(S)}_{\beta}(t_2) | b\rangle \\
\gamma_{ab,cd} (t) &\equiv& \sum_{\alpha, \beta} \int_0^t \diff t_1 \int_0^t \diff t_2 C_{\alpha\beta}(t_1, t_2) \langle a | O^{(S)}_{\beta}(t_2) | b\rangle \langle c | O^{(S)}_{\alpha}(t_1) | d \rangle ^* \\
C_{\alpha\beta}(t_1, t_2) &\equiv& \Tr_E(O^{(E)}_{\alpha}(t_1)O^{(E)}_{\beta}(t_2)\rho_E)
\,,
\eea
where $\{ |a\rangle \}$ forms a complete set of states in the Hilbert space of the subsystem. The Lindblad equation has two non-trivial terms: First, the term proportional to $\sigma_{ab}$ corresponds to a unitary evolution induced by the interaction with the environment, in addition to the usual time evolution driven by the subsystem Hamiltonian. Second, the term proportional to $\gamma_{ab,cd}$ generates a non-unitary evolution through which the subsystem dissipates and loses coherence.

In our case, the subsystem is an energetic jet propagating through the environment which is a QGP in thermal equilibrium with a temperature $T$. For current heavy ion collision experiments, the measured jet energy $Q$ is much larger than the highest temperature achieved in the collision: $Q \gg T$. The jet starts out as a single, highly virtual parton which then showers even in vacuum. Now inside the QGP, the parton shower is modified by its interaction with the medium. In the simple case of forward scattering, the energetic parton exchanges a small amount (small compared with the jet energy) of transverse momentum with the medium. Other physical processes are also possible: A virtual parton can radiate off soft on-shell partons with energy $\sim T$ which then become part of the medium (the wake of a jet). The virtuality of the parton may come from the initial production as in vacuum parton shower, or be developed by a sequence of soft kicks from the medium. In the latter case, the radiation is medium-induced. In this paper, we will take a first step towards understanding the full consequences of the subsystem-environment interaction. To that end, we will ignore the subsystem evolution in vacuum (which is already well understood). We will concentrate on the jet evolution induced by the forward scattering, i.e., the transverse momentum broadening of a jet. We will leave the inclusion of medium-induced radiation into our framework to future studies.

In order to describe the subsystem evolution in terms of the Lindblad equation, we need to know all the three Hamiltonians that are involved:
\begin{itemize}
\item
The subsystem Hamiltonian is the same as the SCET Hamiltonian for collinear particles in vacuum. Usually in vacuum the modes that appear in the SCET Lagrangian are determined by the measurement performed on the jet. Here both the measurements and the medium scales will determine the modes in SCET.
\item
The environment Hamiltonian describes the thermal QGP with the temperature $T$. The sources of the Glauber exchange from the medium are soft modes that scale as $Q(\lambda,\lambda,\lambda)$. 
\item
The interaction Hamiltonian describes the effective interaction between the collinear particles (subsystem) and the soft partons in the QGP (environment). For the near forward scattering region, the interaction happens via Glauber exchanges. The Glauber mode scales as $Q(\lambda, \lambda^2, \lambda)$ with the power counting parameter $\lambda \sim \theta$. The interaction between collinear and soft modes mediated via Glaubers is described in Section~\ref{sec:SCET}.
\end{itemize}

Finally, to apply the Lindblad equation in our study, we need the one-to-one correspondence between the SCET operators and the general operators used in the Lindblad equation. Here we list them:
\bea
O^{(S)}_\alpha(t) &\to& \ml{O}_n^{qA}(t,{\bs x})\,,\  \ml{O}_n^{gA}(t,{\bs x}) \\
\label{eqn:O_E}
O^{(E)}_\alpha(t) &\to& \frac{1}{\ml{P}^2_\perp} \ml{O}_s^{q_nA}(t,{\bs x})\,,\  \frac{1}{\ml{P}^2_\perp}\ml{O}_s^{g_nA}(t,{\bs x}) \\
\alpha &\to& q/g\,,\ {\bs x}\,,\ A 
\eea
where $q/g$ denotes the quark/gluon operator, $A$ is the color index and ${\bs x}$ is the spatial coordinate of the fields.

Now we will apply the Lindblad equation to study the interaction between a collinear quark and a soft quark via the Glauber exchange.

\subsection{Unitary Evolution Part}

We first compute the piece for the unitary evolution of the subsystem density matrix induced by its interaction with the thermal environment
\bea 
\rho_S(t)= \rho_S(0)-i \sum_{a,b} \sigma_{ab}(t)\big[L_{ab}, \rho_S(0)\big] + \cdots \,,
\eea
where non-unitary Lindblad terms are omitted for the moment.
To evaluate the expression, we first write
\bea
\text{sgn}(t_1-t_2) = \Theta(t_1-t_2)-\Theta(t_2-t_1) \,.
\eea
At the same time, we note that
\bea
 &&\sum_{a,b}\bigg( -\frac{i}{2}\sum_{\alpha, \beta}\int_0^t \diff t_1 \int_0^t \diff t_2 C_{\alpha \beta}(t_1, t_2)\Theta(t_1-t_2)\langle a|O^{(S)}_{\alpha}(t_1) O^{(S)}_{\beta}(t_2)|b\rangle L_{ab} \bigg)^{\dagger} \nn\\
&=& \sum_{a,b} -\frac{i}{2}\sum_{\alpha, \beta}\int_0^t \diff t_1 \int_0^t \diff t_2 C_{\alpha \beta}(t_1, t_2)\big(- \Theta(t_2-t_1) \big) \langle a|O^{(S)}_{\alpha}(t_1) O^{(S)}_{\beta}(t_2)|b\rangle L_{ab} \,,\ \ 
\eea
where we have swapped $a$ and $b$, $\alpha$ and $\beta$, $t_1$ and $t_2$ in the last line. Then we can simply write 
\bea
 && \sum_{a,b}\sigma_{ab}L_{ab} \\
&=& \sum_{a,b} -\frac{i}{2}\sum_{\alpha, \beta}\int_0^t \diff t_1 \int_0^t \diff t_2 \langle O^{(E)}_{\alpha}(t_1)O^{(E)}_{\beta}(t_2) \rangle _T  \Theta(t_1-t_2) \langle a| O^{(S)}_{\alpha}(t_1)O^{(S)}_{\beta}(t_2)|b\rangle L_{ab} + \ma{h.c.} \,, \nn 
\eea
where we have replaced the environment correlator
$C_{\alpha \beta}(t_1, t_2) \equiv \Tr_E(O^{(E)}_{\alpha}(t_1)O^{(E)}_{\beta}(t_2)\rho_E) $ with the finite temperature Green's function $\langle O^{(E)}_{\alpha}(t_1) O^{(E)}_{\beta}(t_2) \rangle_T $, 
since the environment density matrix is assumed to be in thermal equilibrium. The subscript $T$ indicates the finite temperature. As we discussed earlier, the indices $\alpha$ and $\beta$ include the spatial coordinates of the field operators. To show the spatial coordinates more explicitly, we rewrite this term as
\bea
\label{eqn:sigmaL}
\sum_{a,b}\sigma_{ab}L_{ab} &=& \sum_{a,b} -\frac{i}{2}\sum_{A, B}\int_0^t \diff t_1 \int_0^t \diff t_2 \int \diff^3 {\bs x}_1 \int \diff^3 {\bs x}_2 \Big[\langle O^{(E)}_{A}(x_1)O^{(E)}_{B}(x_2) \rangle_T \Big]\nn\\
&\times&\Big[\Theta(t_1-t_2)\langle a|O^{(S)}_{A}(x_1)O^{(S)}_{B}(x_2)|b\rangle L_{ab}\Big]+ \ma{h.c.} \,,
\eea
in which $x_i = (t_i,{\bs x}_i)$ and now the indexes $A$ and $B$ are just color indexes and no longer include the spatial coordinates.
If we assume the thermal bath is homogeneous in space and time, we have
\bea
 \langle O^{(E)}_{A}(x_1)O^{(E)}_{B}(x_2) \rangle_T = \langle O^{(E)}_{A}(x_1-x_2) O^{(E)}_{B}(0) \rangle_T \equiv \int \frac{\diff^4k}{(2\pi)^4} e^{-i k\cdot(x_1-x_2)} D_>^{AB}(k) \,, \ \ 
\eea
where $O^{(E)}$ is given by the soft operators (\ref{eqn:O_E}) that are dressed with soft Wilson lines. Here we introduced the finite temperature Wightman function in momentum space $D_>^{AB}(k)$.
We will evaluate this finite temperature correlator for soft quark operators in the imaginary time formalism. Details are provided in Appendix~\ref{sec:T_Correlator}.


\subsubsection{Subsystem Transition}
Now we discuss the computation of the other piece in Eq.~(\ref{eqn:sigmaL}): $ \langle a|O^{(S)}_{\alpha}(x_1)O^{(S)}_{\beta}(x_2)|b\rangle $. Since the purpose of the current paper is to study the transverse momentum kicks, in which splittings are not taken into account, all the relevant states in the subsystem are one particle states. They can be specified by their momentum (we neglect the quark mass), color and spin. Spin does not flip in the Glauber exchange process, which is explained in Appendix~\ref{sec:rate_kernel}. We will average (sum over) the colors of the incoming (outgoing) states. So we will focus on the momentum and use $|p\rangle$ to label the subsystem state $|a\rangle$. In the following, all the abstract state labels $a,b,\cdots$ will be replaced with the momenta $p_1,p_2,\cdots$ and the summation over $a$ will be replaced with an integration over the momentum.

We can then evaluate the correlator of the subsystem (collinear quarks) operators as follows. For two collinear momenta $p_1$ and $p_2$, we define the relevant transition between them as
\bea
J_S \equiv \Theta(t_1-t_2)\langle p_1| O^{(S)}_{\alpha}(t_1)O^{(S)}_{\beta}(t_2)|p_2\rangle \,,
\eea
where the indices $\alpha$ and $\beta$ include both color and the spatial coordinates.
Inserting a complete set of one particle states leads to
\bea
J_S = \Theta(t_1-t_2)\langle p_1|O^{(S)}_{\alpha}(t_1)\int \frac{\diff ^3{\bs q}}{(2\pi)^3 2E_q }|q\rangle\langle q| O^{(S)}_{\beta}(t_2)|p_2\rangle \,,
\eea
where $q^0 = E_q = |{\bs q}|$.
This now represents a $t$ channel process which we are interested in describing, since we focus on a collinear quark.
Since we are concerned with a jet initiated by an energetic quark, we use the appropriate SCET interaction operators in Eq.~(\ref{EFTOp})
\bea
J_S &=& \Theta(t_1-t_2)\int \frac{\diff^3q}{ (2\pi)^3 2E_q}
\Big[\bar u_n(p_1)e^{i(p_1-q)\cdot x_1}\frac{\slashed{\bar n}}{2} T^A u_n(q)\bar u_n(q)\frac{\slashed{\bar n}}{2} T^B u_n(p_2)e^{-i(p_2-q)\cdot x_2}\Big]\nn\\
&=&  \Theta(t_1-t_2)\int \frac{\diff^3q}{ (2\pi)^3 2E_q}\Big[\bar u_n(p_1)e^{i(p_1-q)\cdot x_1}\frac{\slashed{\bar n}}{2} T^A \bar n \cdot q \frac{\slashed n}{2}\frac{\slashed{\bar n}}{2} T^B u_n(p_2)e^{-i(p_2-q)\cdot x_2}\Big] \,.
\eea
This can be rewritten as 
\bea
 J_S= -\frac{1}{2\pi i}\int \frac{\diff^4 q}{ (2\pi)^3 2E_q} \frac{1}{ (q^0-E_q+i\epsilon)}\Big[\bar u_n(p_1)e^{i(p_1-q)\cdot x_1} \frac{\slashed{\bar n}}{2} \bar n \cdot q T^A T^B u_n(p_2)e^{-i(p_2-q)\cdot x_2}\Big] \,,\nn\\
\eea
where now $q^0$ is not constrained.

\subsubsection{Final Expression}
We can now put all the pieces together for the driving term of the unitary evolution
\bea
\sum_{a,b}\sigma_{ab}L_{ab} &=& -\frac{i}{2}\int \widetilde{\diff p_1} \int \widetilde{\diff p_2} \sum_{A, B}\int_0^t \diff t_1 \int_0^t \diff t_2 \int \diff^3 {\bs x}_1 \int \diff^3 {\bs x}_2 \frac{1}{N_c} D_>^{A B}(x_1-x_2) J^{AB}_S(x_1,x_2)\nn\\
&=& \frac{i}{2}\int \widetilde{\diff p_1} \int \widetilde{\diff p_2} \sum_{A, B}\int_0^t \diff t_1 \int_0^t \diff t_2 \int \diff^3 {\bs x}_1 \int \diff^3 {\bs x}_2\nn\\
&\times&\int \frac{\diff^4k}{(2\pi)^4} e^{-i k\cdot(x_1-x_2)} D_>^{A B}(k)\frac{1}{2\pi i} \int \frac{\diff^4 q}{(2\pi)^3 2E_q } \frac{1}{ (q^0-E_q+i\epsilon)} \nn\\
&\times&\frac{1}{N_c}\Big[\bar u_n(p_1)e^{i(p_1-q)\cdot x_1}\frac{\slashed{\bar n}}{2}\bar n \cdot q  T^{A} T^{B}u_n(p_2)e^{-i(p_2-q)\cdot x_2}\Big] |p_1\rangle \langle p_2| + \ma{h.c.}\,,\ \ \ \ 
\eea
where the $1/N_c$ factor comes from the average of the color of the incoming states. We have introduced the shorthand notation
\bea
  \widetilde{\diff p_i} = \frac{\diff^3 {\bs p}_i}{ (2\pi)^3 2E_{p_i} }\,.
\eea
Integrating over ${\bs x}_1$ and ${\bs x}_2$ gives two delta functions for momentum conservation: $\delta^3({\bs p}_1 - {\bs q} -{\bs k})$ and $\delta^3({\bs p}_2 - {\bs q} -{\bs k})$. From this we can conclude ${\bs p}_1 = {\bs p}_2 \equiv {\bs p}$. Since the external one particle states $|p_i\rangle$ ($i=1,2$) are on-shell, we further conclude $E_{p_1} = E_{p_2} \equiv E_p$. Then the relevant time integrals in the limit $t\rightarrow \infty$ become
\bea
\label{eqn:markov_time_int}
\int_0^t \diff t_1 \int_0^t \diff t_2 e^{i\omega t_1}e^{-i\omega t_2} \xrightarrow{t\to\infty} 2\pi t\delta(\omega)
\eea 
where $\omega = E_p - k^0 - q^0$ is the common energy conservation condition that the integral over $t_i$ yields. The $t\to\infty$ limit is called the Markovian approximation, which is valid when the subsystem relaxation time is much bigger than the environment correlation time. Physically it means during a typical evolution time of the subsystem, the environment loses all memory of the subsystem. Mathematically, it occurs when the environment correlator $C_{\alpha \beta}(t_1, t_2)$ dies off quickly enough at large $t_i$ so the contribution to the time integral from the large time region is negligible. Markovian approximation is generally true when the subsystem is weakly coupled to a thermal bath. The argument is as follows: The typical energy scale of the thermal bath is the temperature. Its inverse gives the typical correlation time of the environment. The subsystem relaxation rate is roughly $\alpha_s T$, which is based on the leading order estimate. The inverse of the relaxation rate gives the relaxation time, which is much bigger than $1/T$ when $\alpha_s$ is small. We will give an explicit estimate of the subsystem relaxation rate in Section~\ref{sect:markovian}.

Finally we obtain
\bea
 \sum_{a,b}\sigma_{ab}L_{ab} &=& \frac{t}{2} \sum_{AB} \int \frac{\widetilde{\diff p}}{2E_p} \int \frac{\diff^4 k}{(2\pi)^4}  \int \frac{\diff^4 q}{(2\pi)^4 2E_q}\frac{1}{ q^0-E_q+i\epsilon} \frac{1}{N_c} \Big[\bar u_n(p)\frac{\slashed{\bar n}}{2}\bar n \cdot q T^{A} T^{B}u_n(p)\Big] \nn\\
 &\times& D_>^{AB}(k) (2\pi)^4\delta^3({\bs p}-{\bs q}-{\bs k})\delta(E_p-k^0-q^0) |p\rangle \langle p|+ \ma{h.c.} \,.
\eea
Since the various momenta in the formula scale in a specific manner, we can simplify the formula further by doing an expansion and keeping terms only at leading power in $\lambda$.

We notice that the unitary evolution driving term is diagonal in the subsystem state space. If the initial state density matrix is a pure state of a single parton $\rho_S(0) = |Q_0\rangle \langle Q_0|$, we find
\bea
 -i\sum_{a,b} \sigma_{ab}(t)\big[L_{ab}, \rho_S(0)\big] = 0 \,.
\eea
So at least at leading order, no correction on the subsystem unitary evolution is generated from its interaction with the medium. In other words, at leading order, the single parton state energy is not corrected by the medium interaction.


\subsection{Non-unitary (Dissipative) Evolution Part}
The other terms in the Lindblad equation lead to a non-unitary or dissipative evolution for the subsystem density matrix:
\bea
 \rho_S(t)= \rho_S(0) + \sum_{a,b,c,d} \gamma_{ab,cd}(t)\Big( L_{ab}\rho_S(0)L_{cd}^{\dagger}-\frac{1}{2}\{L_{cd}^{\dagger}L_{ab}, \rho_S(0)\}\Big)+... \,,
\eea
where higher order terms are neglected.
Even though the evolution is non-unitary, it preserves the properties of a valid density matrix, i.e., hermiticity, positivity and unity trace (the total density of all states is conserved).
In this case the correlator in the environment does not have any time ordering. We can simply relate it to one of the Wightman functions using translational invariance of the thermal bath, 
\bea
\Tr_E\big( O^{(E)}_{\alpha}(x_1)O^{(E)}_{\beta}(x_2)\rho_E \big) &=& \langle  O^{(E)}_{A}(x_1-x_2)O^{(E)}_{B}(0) \rangle_T \nn\\
&=& \int \frac{\diff^4k}{(2\pi)^4} e^{-ik \cdot (x_1-x_2)}D_{>}^{AB}(k) \,,
\eea
which can be related to the other Wightman function as well as the spectral function. For our bosonic environment operators (\ref{eqn:O_E}),
\bea
D_{<}^{AB}(k) = e^{-k_0/T }D_{>}^{AB}(k) = n_B(k^0)\rho^{AB}(k) \,,
\eea
where $n_B$ is the Bose-Einstein distribution and $\rho^{AB}(k)$ is the spectral function.
Details for the weak coupling computation of the Wightman functions are given in Appendix~\ref{sec:T_Correlator}.

Now we compute the matrix element of the subsystem operator:
\bea
 \langle a| O^{(S)}_{\beta}(t_2)|b\rangle \langle c|O^{(S)}_{\alpha}(t_1)|d\rangle^* = \langle d| O^{(S)}_{\alpha}(t_1)|c\rangle  \langle a| O^{(S)}_{\beta}(t_2)|b\rangle \,.
\eea
Since we focus on one particle quark state with a collinear momentum, we can use the collinear momentum to label the states. As discussed in the unitary evolution, spin does not flip in the transition and we will average over the initial state colors and sum over the final state colors. We choose $|a\rangle = |p_1\rangle$, $|b\rangle = |p_2\rangle$, $|c\rangle = |p_3\rangle$ and $|d\rangle = |p_4\rangle$. Then at leading order we have,
\bea
 &&\langle p_4|O^{(S)}_{\alpha}(t_1)|p_3\rangle  \langle p_1|O^{(S)}_{\beta}(t_2)|p_2\rangle \nn \\
 &=& \bigg[\bar u_n(p_4)\frac{\slashed{\bar n}}{2}T^A u_n(p_3)\bar u_n(p_1)\frac{\slashed{\bar n}}{2}T^B u_n(p_2)\bigg]e^{-i(p_3-p_4)\cdot x_1 }e^{-i(p_2-p_1) \cdot x_2} \,.
\eea
We can now separately evaluate each term in our expression for non-unitary evolution. We first compute
\bea
&&\sum_{a,b,c,d} \gamma_{ab,cd}(t)L_{cd}^{\dagger} L_{ab}= \prod_{i=1}^4\int \widetilde{\diff p_i} \int_0^t \diff t_1 \int_0^t \diff t_2 \int \diff^3{\bs x}_1 \int \diff^3 {\bs x}_2 \int\frac{\diff^4k}{(2\pi)^4} e^{-ik \cdot (x_1-x_2)}\nn\\
&\times& D_{>}^{AB}(k) \frac{1}{N_c} \bigg[\bar u_n(p_4)\frac{\slashed{\bar n}}{2}T^{A} u_n(p_3)\bar u_n(p_1)\frac{\slashed{\bar n}}{2}T^{B}u_n(p_2)\bigg]e^{-i(p_3-p_4)\cdot x_1}e^{-i(p_2-p_1)\cdot x_2} L_{34}^{\dagger}L_{12} \,. \ \ \ \ \ \ \ \ 
\eea
Since $L_{34}^{\dagger}L_{12}= |p_4\rangle \langle p_3 |p_1 \rangle \langle p_2| = 2E_{p_3}(2\pi)^3 \delta^3({\bs p}_3 - {\bs p}_1) |p_4\rangle \langle p_2|$, we can use this to eliminate the integral over one of the momenta. The integrals over ${\bs x}_1$ and ${\bs x}_2$ then set ${\bs p}_2={\bs p}_4$. We further obtain $E_{p_2} = E_{p_4}$ and $E_{p_1} = E_{p_3}$ because of the on-shell particles. Then we can apply the same trick to the time integrals as we did for the unitary evolution in the previous subsection. The two time integrals will lead to one delta function for energy conservation, multiplied by the time length $t$ in the Markovian approximation. When the dust settles, we are left with 
\bea
\sum_{a,b,c,d} \gamma_{ab,cd}(t) L_{cd}^{\dagger} L_{ab} &=& t\int \frac{\widetilde{\diff p}}{2E_p} \int \widetilde{\diff q} \int \frac{\diff^4k}{(2\pi)^4} D_{>}^{AB}(k) \frac{1}{N_c} \bigg[\bar u_n(p)\frac{\slashed{\bar n}}{2}   \bar n \cdot q  T^{A} T^{B}u_n(p)\bigg]\nn\\
&\times&(2\pi)^4\delta^3({\bs p}- {\bs k}-{\bs q})\delta(E_p - k^0 - E_q) |p\rangle \langle p|\,,
\eea 
where $E_p = |{\bs p}|$ and $E_q =|{\bs q}|$. The dummy color indexes $A$ and $B$ are summed over implicitly. This term appears as
\be
-\frac{1}{2}\sum_{a,b,c,d}\gamma_{ab,cd}(t) \{ L_{cd}^{\dagger} L_{ab}, \rho_S(0) \} \,,
\ee
in the Lindblad equation. 
When the initial subsystem density matrix is $ |Q_0\rangle \langle Q_0|$, it gives back a projection onto $ |Q_0\rangle \langle Q_0|$. Then the anti-commutator gives the same result, which cancels the factor of one half. Due to the negative sign, this term in the Lindblad equation represents the loss in the probability of staying in the state $|Q_0\rangle \langle Q_0|$.

We have one more term in the Lindblad equation:
\bea
&& \sum_{a,b,c,d} \gamma_{ab,cd}(t) L_{ab}\rho_s(0)L_{cd}^{\dagger} =\prod_{i=1}^4\int \widetilde{\diff p_i} \int_0^t \diff t_1 \int_0^t \diff t_2 \int \diff^3 {\bs x}_1 \int \diff^3 {\bs x}_2 \int \frac{\diff^4k}{(2\pi)^4}e^{-i k \cdot (x_1-x_2)}\nn\\
&\times&D_{>}^{AB}(k) \frac{1}{N_c} \bigg[\bar u_n(p_4)\frac{\slashed{\bar n}}{2}T^{A}u_n(p_3)\bar u_n(p_1)\frac{\slashed{\bar n}}{2}T^{B}u_n(p_2)\bigg]e^{-i(p_3-p_4)\cdot x_1}e^{-i(p_2-p_1) \cdot x_2}L_{12}\rho(0)L_{34}^{\dagger} \,, \nn\\
\eea
where the Lindblad operator is defined by $L_{ij} = |p_i\rangle \langle p_j |$. In the Markovian approximation $t\to\infty$\footnote{Rigorously speaking, one must first show the two frequencies associated with time in the exponents are equal and then apply Eq.~(\ref{eqn:markov_time_int}) to obtain the energy conservation delta function multiplied by the time length. Here we write down two delta functions for energy conservation without the factor of time length $t$. Later we will show the two energy conservation functions are the same for our case here, which allows us to write one of them as the time length.},
\bea
&& \sum_{a,b,c,d} \gamma_{ab,cd}(t) L_{ab}\rho_s(0)L_{cd}^{\dagger} =\prod_{i=1}^4\int \widetilde{\diff p_i} \int \frac{\diff^4k}{(2\pi)^4}  \frac{1}{N_c}\bigg[\bar u_n(p_4)\frac{\slashed{\bar n}}{2}T^{A}u_n(p_3)\bar u_n(p_1)\frac{\slashed{\bar n}}{2}T^{B}u_n(p_2)\bigg] \nn\\
&\times& D_{>}^{AB}(k) (2\pi)^8\delta^4(p_3+k-p_4)\delta^4(p_1+k-p_2) |p_1\rangle\langle p_2|\rho(0)|p_4\rangle \langle p_3| \,.
\eea
Further simplification depends on the initial subsystem density matrix $\rho_S(0)$ and the final measurement operator applied.



\section{Transverse Momentum Broadening}
\label{sec:Transverse}
We now apply the Lindblad equation to compute specific observables on the subsystem density matrix. For an observable associated with a subsystem operator $M$, the measurement result at time $t$ is defined by the expectation value
\bea
\langle M \rangle(t) = \Tr_S \big( M \rho_S(t) \big)  \,.
\eea
In our case, the Fock state of the subsystem consists of only one particle states, labelled by a collinear momentum. So we can write
\be
\langle M \rangle(t) = \int \widetilde{\diff p} \int \widetilde{\diff q} \,\langle p |M| q \rangle \langle q |\rho_S(t) | p \rangle \,.
\ee
In the future, when we include splitting in the study, we will need to include all possible states in the Fock space such as a two particle state.

An interesting observable is the final transverse momentum distribution of particles within a jet (with respect to the jet axis). If we focus on the leading parton in the jet, the final transverse momentum vanishes in vacuum. But in the medium, due to the interactions with the medium, the leading parton can develop a non-zero final transverse momentum. In other words, the transverse momentum distribution is broadened by the interactions with the medium. To obtain the final momentum distribution, we set the measurement operator to be a projection operator $P_Q = |Q\rangle \langle Q|$ for some momentum of $Q$. Then we need to compute
\be
\langle P_Q \rangle(t) = \langle Q |\rho_S(t) |Q\rangle \,.
\ee
We will derive an evolution equation for the observable $\langle P_Q \rangle(t)$ that will correspond to a linear differential equation in time. By solving it, we can resum various effects. The resummation can be done given that the Markovian approximation holds. To obtain the measurement results, our job now is to compute the time evolution of the diagonal piece of the subsystem density matrix.

We first put all non-vanishing leading order pieces derived in the previous section together
\bea
 \rho_S(t)&=&\rho_S(0) -\frac{t}{2}\int \frac{\widetilde{\diff p}}{2E_p} \int \widetilde{\diff q} \int \frac{\diff^4k}{(2\pi)^4} D_{>}^{AB}(k) 
 \frac{1}{N_c}\bigg[\bar u_n(p)\frac{\slashed{\bar n}}{2}   \bar n \cdot q  T^{A} T^{B}u_n(p)\bigg]\nn\\
&\times&(2\pi)^4\delta^4( p- k - q) \big\{ |p\rangle \langle p| ,\rho_S(0) \big\} \nn\\
&+&\prod_{i=1}^4\int \widetilde{\diff p_i} \int \frac{\diff^4k}{(2\pi)^4} D_{>}^{AB}(k) \frac{1}{N_c} \bigg[\bar u_n(p_4)\frac{\slashed{\bar n}}{2}T^{A}u_n(p_3)\bar u_n(p_1)\frac{\slashed{\bar n}}{2}T^{B}u(p_2)\bigg]\nn\\
&\times&(2\pi)^8\delta^4(p_3+k-p_4)\delta^4(p_1+k-p_2) |p_1\rangle\langle p_2|\rho_S(0)|p_4\rangle \langle p_3| \,,
\eea
where we neglect the medium-induced unitary evolution since later we will assume the initial density matrix is a single parton with a given momentum.
Sandwiching between $\langle Q|$ and $|Q\rangle $ leads to
\bea
 && \langle Q|\rho_S(t)|Q \rangle = \langle Q|\rho_S(0)|Q \rangle \nn\\
 &-& t \int \frac{\widetilde{\diff q}}{2E_Q} \int \frac{\diff^4k}{(2\pi)^4} D_{>}^{AB}(k) \frac{1}{N_c} \bigg[\bar u_n(Q)\frac{\slashed{\bar n}}{2}\bar n \cdot q T^{A} T^B u_n(Q)\bigg](2\pi)^4\delta^4(Q-k-q)\langle Q|\rho_S(0)|Q\rangle \nn\\
&+& t \int \frac{\widetilde{\diff q}}{2E_q} \int \frac{\diff^4k}{(2\pi)^4} D_{>}^{AB}(k) \frac{1}{N_c} \bigg[\bar u_n(q)\frac{\slashed{\bar n}}{2}T^A u_n(Q)\bar u_n(Q)\frac{\slashed{\bar n}}{2}T^B u(q)\bigg] \nn \\ 
&\times& (2\pi)^4\delta^4(Q+k-q)\langle q|\rho_S(0)|q\rangle \,,
\eea 
which schematically can be written as
\bea
\label{eqn:master0}
 \langle Q|\rho_S(t)|Q \rangle =\langle Q|\rho_S(0)|Q \rangle  - t R(Q)\langle Q|\rho_S(0)|Q\rangle +t \int \widetilde{\diff q} K(Q,q) \langle q|\rho_S(0)|q\rangle \,,
\eea
where we defined the dissipation rate $R(Q)$ and the fluctuation kernel $K(Q,q)$
\bea
\label{eqn:rate}
R(Q) &=& \int\frac{\widetilde{\diff q}}{2E_Q} \int \frac{\diff^4k}{(2\pi)^4} D_{>}^{AB}(k) \frac{1}{N_c} \bigg[\bar u_n(Q)\frac{\slashed{\bar n}}{2}\bar n \cdot q T^{A} T^B u_n(Q)\bigg](2\pi)^4\delta^4(Q-k-q) \\
\label{eqn:kernel}
K(Q,q) &=& \frac{1}{2E_q} \int \frac{\diff^4k}{(2\pi)^4} D_{>}^{AB}(k) \frac{1}{N_c} \bigg[\bar u_n(q)\frac{\slashed{\bar n}}{2}T^A u_n(Q)\bar u_n(Q)\frac{\slashed{\bar n}}{2}T^B u(q)\bigg](2\pi)^4\delta^4(Q+k-q) \,.\nn\\
\eea

\subsection{Markovian Approximation}
\label{sect:markovian}
To convert Eq.~(\ref{eqn:master0}) into a different equation in time, we will use the Markovian approximation again. We will move the first term on the right hand side to the left hand side, divide the equation by $t$ and take the limit $t\to0$. Then we infer the master equation for the probability of being in a specific momentum state $P(Q,t) \equiv \langle Q|\rho_S(t)|Q \rangle$
\bea
\label{eqn:master}
\partial_t P(Q,t) = -R(Q)P(Q,t) + \int \widetilde{\diff q} K(Q,q) P(q,t) \,.
\eea
One should be cautious here because previously we have taken $t\to\infty$ when computing the time integral to obtain delta functions for energy conservation. These two seemingly contradictory limits are compatible with each other in the Markovian limit. The Markovian approximation is valid if the environment correlation time is much smaller than the subsystem relaxation time. When we take $t\to0$, we are thinking of the time length as a typical subsystem relaxation time. This time length is still much larger than the environment correlation time. What seems to be a short time for the subsystem is actually very long for the environment. The master equation (\ref{eqn:master}) is coarse-grained. Physically, the environment has lost any information about the subsystem before it interacts again with it. In this way, at each interaction point, the environment has no memory about the past history of the subsystem.

Using Eq.~(\ref{eqn:master}) derived above, we can check the validity of the Markovian approximation. The dissipation rate $R(Q)$ is associated with a typical time scale of the subsystem relaxation $1/R(Q)$, which is originated from the Glauber exchange between a collinear parton and a soft parton from the medium. On the other hand, the typical time scale for the environment decoherence is of the order of $1/T$, with $T$ being the temperature of the thermal bath. So for the validity of the Markovian approximation, we require
\bea
\frac{1}{T} \ll \frac{1}{R(Q)} \,.
\eea
From Eq.~(\ref{eqn:hatR}), which will be explained in Section~\ref{sect:num}, we can estimate the relaxation rate as
\bea
R(Q)\sim T\alpha_s^2 \int_0^{\infty} \frac{|\hat{k}_{\perp}|\diff |\hat{k}_{\perp}|}{(|\hat{k}_{\perp}|^2+m_D^2/T^2)^2}\sim \frac{T^3\alpha_s^2}{m_D^2}\sim T\alpha_s \,,
\eea
so the Markovian approximation is valid in the weak coupling limit, where we have used the fact that $m_D \sim gT$.

The structure of the master equation (\ref{eqn:master}) is simple: The first term on the right hand side is a loss term for the state $|Q\rangle$. The probability of being in the state $|Q\rangle$ decreases with time because it may transition to other momentum states due to the Glauber exchange with the medium. The last term is a gain term for the state $|Q\rangle$. It originates again from the Glauber exchange. States with other momenta, say $q$, can turn into the state $|Q\rangle$ by exchanging momentum with the medium.

\subsection{Solution to Master Equation}
Before we show the solution to the master equation, we want to elucidate our notations. We will use $r_\perp$ and $k_\perp$ to label the Minkowski transverse vectors while ${\bs r}_\perp$ and ${\bs k}_\perp$ to label the Euclidean transverse vectors. For the magnitude, we will use notations such as $|r_\perp|$ and $|k_\perp|$.

We can rewrite the master equation $(\ref{eqn:master})$ in a suggestive form (see Appendix~\ref{sec:rate_kernel} for the explanation)
\bea
\label{eqn:master_sol}
\partial_t P\bigg(\Big[Q^-, \frac{|Q_{\perp}|^2}{Q^-},Q_{\perp} \Big],t \bigg) &=& -R(Q) P\bigg(\Big[Q^-, \frac{|Q_{\perp}|^2}{Q^-},Q_{\perp} \Big],t \bigg) \nn\\
&+& \int \diff^2 k_{\perp} K(Q, k_{\perp}) P\bigg( \Big[Q^-,\frac{|Q_{\perp}+k_{\perp}|^2}{Q^-},Q_{\perp}+k_{\perp} \Big],t \bigg)\,,\ \ \ \ \ 
\eea
where we abused the notation: $K(Q, k_{\perp})$ here includes both $K(Q,q)$ in Eq.~(\ref{eqn:kernel}) and some integration from the last term of Eq.~(\ref{eqn:master0}).
Due to the expansion based on our power counting, we see that the second term is a convolution only in the $\perp$ direction. Hence the obvious way to solve this equation is to move to the space of impact parameter. Defining
\bea
 P\bigg( \Big[Q^-,\frac{|q_{\perp}|^2}{Q^-},q_{\perp} \Big],t \bigg) &\equiv & \int \diff^2 r_{\perp}e^{-i {\bs r}_{\perp}\cdot {\bs q}_{\perp}} \widetilde{P}(Q^-,r_{\perp},t) \\
\label{eqn:kernel_momentum}
K(Q, k_{\perp} ) &\equiv&  \int \diff^2 s_{\perp}e^{-i {\bs s}_{\perp}\cdot {\bs k}_{\perp}} \widetilde{K}(Q, s_{\perp}) \,,
\eea 
we can simplify Eq.~(\ref{eqn:master_sol}) as 
\bea
&& \partial_t \int \diff^2 r_{\perp} e^{-i {\bs r}_{\perp}\cdot {\bs Q}_{\perp}} \widetilde{P} (Q^-,r_{\perp},t) = -R(Q) \int \diff^2 r_{\perp} e^{-i {\bs r}_{\perp}\cdot {\bs Q}_{\perp}} \widetilde{P}(Q^-,r_{\perp},t)\nn\\
&+& \int \diff^2 k_{\perp} \int \diff^2 s_{\perp}e^{-i {\bs s}_{\perp}\cdot {\bs k}_{\perp}} \widetilde{K}(Q, s_{\perp})
\int \diff^2 r_{\perp}e^{-i {\bs r}_{\perp}\cdot ({\bs Q}_{\perp} + {\bs k}_\perp) } \widetilde{P}(Q^-,r_{\perp},t) \nn\\
&=& -R(Q) \int \diff^2 r_{\perp} e^{-i {\bs r}_{\perp}\cdot {\bs Q}_{\perp}} \widetilde{P}(Q^-,r_{\perp},t)
+ \int \diff^2 r_{\perp} e^{-i {\bs r}_{\perp}\cdot {\bs Q}_{\perp}}
\widetilde{K}(Q, -r_{\perp})
\widetilde{P}(Q^-,r_{\perp},t)\,.\ \ \ \ \ \ \ 
\eea
We now obtain a simpler equation 
\bea
 \partial_t \widetilde{P}(Q^-,r_{\perp},t) &=& - R(Q) \widetilde{P}(Q^-,r_{\perp},t) + \widetilde{K}(Q, -r_{\perp})
\widetilde{P}(Q^-,r_{\perp},t) \nn\\
&=&  \big[ -R(Q) + \widetilde{K}(Q, -r_{\perp}) \big] \widetilde{P}(Q^-,r_{\perp},t) \,,
\eea
which suggests a solution of the form
\bea
\widetilde{P}(Q^-,r_{\perp},t) = e^{ \big[- R(Q) + \widetilde{K}(Q, -r_{\perp}) \big] t} \widetilde{P}(Q^-,r_{\perp},t=0) \,.
\eea
Then we can write out the final solution of our measurement results at time $t$ as
\bea
P(Q^-,Q_{\perp},t) &=& \int \diff^2 r_{\perp} e^{-i {\bs r}_{\perp}\cdot {\bs Q}_{\perp}}e^{ \big[- R(Q) + \widetilde{K}(Q, -r_{\perp}) \big] t} \widetilde{P}(Q^-,r_{\perp},t=0) \,. 
\eea
 We will examine the distribution in transverse momentum, starting out with a collinear quark that has zero transverse momentum. Then our initial condition for the subsystem density matrix is
\bea
P\bigg( \Big[Q^-,\frac{|Q_\perp|^2}{Q^-}, Q_\perp \Big], t=0 \bigg) &=& f(Q^-) \delta^2({\bs Q}_\perp)\\
\widetilde{P}(Q^-,r_{\perp},t=0) &=& \frac{f(Q^-)}{(2\pi)^2} \,,
\eea
in which $f(Q^-)$ is the overall normalization of the initial density. If we just focus on the transverse momentum distribution, we can set $f(Q^-)=(2\pi)^2$. Our solution then becomes 
\bea
\label{eqn:sol_to_master621}
 P(Q^-,Q_{\perp},t) = \frac{f(Q^-)}{(2\pi)^2} \int \diff^2 r_{\perp}e^{-i {\bs r}_{\perp} \cdot{\bs Q}_{\perp}   } e^{ \big[- R(Q) + \widetilde{K}(Q, -r_{\perp}) \big] t} \,.
\eea
For the distribution at any time $t$, we only need to evaluate the $T$ and $Q_{\perp}$ dependent Sudakov factor $S(Q,{\bs r}_\perp) \equiv -R(Q) + \widetilde{K}(Q, -r_{\perp})$. Using the results from Appendix~\ref{sec:rate_kernel}, we find for a collinear quark scattered off soft quarks of the medium
\bea
S(Q,{\bs r}_\perp) &=& - \frac{C_F}{2}  \int \frac{\diff^2 k_{\perp} \diff k^-}{(2\pi)^3} \Big[ 1 - e^{-i{\bs k}_\perp \cdot {\bs r}_\perp} \Big] D_>(k^-,k_{\perp})  \\ 
&=&\frac{2\alpha_s^2N_fC_FT_F}{\pi^3} \int \frac{|k_\perp|\diff |k_\perp|}{|k_\perp|^4}
\diff\phi_k \Big[ e^{-i| k_\perp| | r_\perp| \cos\phi_k} -1 \Big]
\int  \diff |p_\perp| \diff p^- \diff \phi\, \frac{|p_\perp|^3}{(p^-)^2}   \nn \\
&\times& n_F\Big(\frac{(p^-)^2+|p_\perp|^2}{2p^-}\Big) \bigg[1-n_F\Big(\frac{ (p^-)^2 (|p_\perp|^2 + |k_\perp|^2 + 2|p_\perp||k_\perp|\cos\phi) +|p_\perp|^4}{2|p_\perp|^2p^-}\Big)  \bigg] \,, \nn
\eea
where $p^->0$, $C_F=\frac{N_c^2-1}{2N_c}$, $T_F = \frac{1}{2}$ and $N_f$ is the number of active quark flavors in the medium.
The angular integration over $\phi_k $ can be done by using the Bessel function of the first kind:
\bea
 J_0(z)=\frac{1}{2\pi}\int_0^{2\pi}\diff \theta e^{iz\cos \theta} \,.
\eea
Then the Sudakov factor can be written as
\bea
\label{eqn:sudakov}
S(Q,{\bs r}_\perp) &=&\frac{4\alpha_s^2N_fC_FT_F}{\pi^2} \int \frac{|k_\perp|\diff |k_\perp|}{|k_\perp|^4}
 \Big[ J_0(|r_{\perp}||k_{\perp}|) -1 \Big]
\int  \diff |p_\perp| \diff p^- \diff \phi\, \frac{|p_\perp|^3}{(p^-)^2}   \\
&\times& n_F\Big(\frac{(p^-)^2+|p_\perp|^2}{2p^-}\Big) \bigg[1-n_F\Big(\frac{ (p^-)^2 (|p_\perp|^2 + |k_\perp|^2 + 2|p_\perp||k_\perp|\cos\phi) +|p_\perp|^4}{2|p_\perp|^2p^-}\Big)  \bigg] \,, \nn
\eea
The integrand over $p^-$ is from $0$ to $\infty$. The integrand is regular as $p^- \to \infty$ or $|p_\perp|\to\infty $ due to the Fermi-Dirac distribution. On the IR side, the seemingly singular point $p^-=0$ is actually regular because the integrand scales as $(p^-)^{-2}e^{-| p_\perp |^2/(2p^-)}$ when $p^-\to0$ for non-vanishing $|p_\perp|$. If $|p_\perp| = 0$ the integrand is vanishing. IR singularity exists when $|k_\perp|\to0$. In this limit, the Bessel function behaves as 
\bea
J_0(|r_{\perp}||k_{\perp}|) = 1 -\frac{|r_{\perp}|^2|k_{\perp}|^2}{4} +\ml{O}\big(|r_{\perp}|^4|k_{\perp}|^4\big) \,.
\eea
So it partially cancels out the singularity of $1/|k_\perp|^3$ at $k_{\perp}=0$. But the Sudakov factor still has a logarithmic singularity. We will discuss this singularity in detail in Section~\ref{sect:ir}. In our numerical studies shown in Section~\ref{sect:num}, we will cut this IR divergence by introducing the Debye screening. 

Another singular behavior can appear in the final solution because our initial density is a delta function in the transverse momentum. To make the initial delta function more explicit in the final solution, we will reorganize our result as follows:
\bea
&&P(Q^-,Q_{\perp},t) = \frac{f(Q^-)}{(2\pi)^2} \int \diff^2 r_{\perp}e^{-i {\bs r}_{\perp} \cdot{\bs Q}_{\perp}   } e^{ \big[- R(Q) + \widetilde{K}(Q, -r_{\perp}) \big] t} \nn \\
&=& \frac{f(Q^-)}{(2\pi)^2} e^{- R(Q)t}  \int \diff^2 r_{\perp}e^{-i {\bs r}_{\perp} \cdot{\bs Q}_{\perp}   } \Big(e^{ \widetilde{K}(Q, -r_{\perp})  t} +1 - 1\Big) \nn\\
\label{eqn:final}
&=& f(Q^-)e^{- R(Q)t}  \delta^2({\bs Q}_\perp) + \frac{f(Q^-)}{(2\pi)^2} \int \diff^2 r_{\perp}e^{-i {\bs r}_{\perp} \cdot{\bs Q}_{\perp}   } \Big( e^{ \big[- R(Q) + \widetilde{K}(Q, -r_{\perp}) \big] t} - e^{-R(Q)t} \Big) \,.\ \ \ \ \ \ 
\eea
The physical meaning of the separation is as follows: The first term describes that the density of the initial state ($Q_\perp=0$) decays over time with the rate $R(Q)$. The second term is the growing of the density of other states ($Q_\perp \neq0$). If we integrate over $Q_\perp$, we will find the total probability is conserved which is to say that the time evolution preserves the trace of our density matrix. 


\subsection{IR Safety}
\label{sect:ir}
We are interested in the physical regime $Q_{\perp} \sim T$ to which only the second term in Eq.~(\ref{eqn:final}) contributes. We define
\bea
\label{eqn:def_G}
  G(Q^-,Q_{\perp},t) \equiv \int \diff^2 r_{\perp}e^{-i {\bs r}_{\perp} \cdot{\bs Q}_{\perp}   } \Big( e^{ - R(Q)t + \widetilde{K}(Q, -r_{\perp}) t } - e^{-R(Q)t} \Big) \,.
\eea
If we expand out the exponent of the right hand side, we have 
\bea
\label{eqn:G_expansion}
 G(Q^-,Q_{\perp},t) &=& \int \diff ^2r_{\perp}e^{-i {\bs r}_{\perp} \cdot{\bs Q}_{\perp}   }  \sum_{n=0}^{\infty} \bigg(\frac{\big[ \widetilde{K}(Q, -r_{\perp}) t - R(Q)t \big]^n}{n!} - \frac{\big[- R(Q)t \big]^n}{n!}\bigg) \,, \ \ \ \ \ 
\eea
where the rate and the kernel have the form of
\bea
\label{eqn:rate_form}
R(Q) &=& \int \frac{\diff^2 k_\perp}{|k_\perp|^4} \ml{W}(k_\perp) \,, \\
\label{eqn:kernel_form}
\widetilde{K}(Q, -r_{\perp}) &=& \int \frac{\diff^2 k_\perp}{|k_\perp|^4} e^{- i{\bs r}_\perp \cdot {\bs k}_\perp} \ml{W}(k_\perp) \,.
\eea
The function $\mathcal{W}(k_{\perp})$ can be read from Eq.~(\ref{eqn:sudakov}) and does not have any singularity at $k_{\perp}=0$. In fact, the function $\mathcal{W}(k_{\perp})$ goes to a constant value at both low and high $k_{\perp}$ (see Fig.~\ref{fig:1}).
For $n=0$ and $1$, we can work out the result easily
\bea
 G^{(0)}(Q^-,Q_{\perp},t) &=& 0 \\
 \label{eqn:G1}
 G^{(1)}(Q^-,Q_{\perp},t) &=& t \int \diff^2 r_{\perp}e^{-i {\bs r}_{\perp}\cdot {\bs Q}_{\perp}} \widetilde{K}(Q, -r_{\perp}) = (2\pi)^2 t \frac{\mathcal{W}(-Q_{\perp})}{|Q_{\perp}|^4}\,.
\eea
For $n=0$ and $1$, the singularity at $k_\perp=0$ does not influence the computation of the integral. The IR singularity is cut-off by the constraint $Q_\perp$ in the final measurements.

We can work out the $n=2$ term in the series to see a non-trivial cancellation of some IR divergences:
\bea
 &&G^{(2)}(Q^-,Q_{\perp},t) =  \frac{t^2}{2}\int \diff^2r_{\perp}e^{-i {\bs r}_{\perp}\cdot {\bs Q}_{\perp}}\Big(\widetilde{K}^2(Q, -r_{\perp}) -2 R(Q) \widetilde{K}(Q, -r_{\perp}) \Big)\nn\\
&=& 2\pi^2 t^2\left(  \int \frac{\diff^2 k_{1\perp}}{|k_{1\perp}|^4}\mathcal{W}(k_{1\perp}) \int \frac{\diff^2 k_{2\perp}}{|k_{2\perp}|^4}\mathcal{W}(k_{2\perp})\delta^2({\bs Q}_{\perp} + {\bs k}_{1\perp} + {\bs k}_{2\perp})- 2 R(Q) \frac{\mathcal{W}(-Q_{\perp})}{|Q_{\perp}|^4}\right) \nn\\
&=& 2\pi^2 t^2 \left(  \int \diff^2 k_{\perp} \frac{\mathcal{W}(k_{\perp})}{|k_{\perp}|^4} \frac{\mathcal{W}(-{ Q}_{\perp}-{ k}_{\perp})}{|{ Q}_{\perp} + { k}_{\perp}|^4}- 2 R(Q) \frac{\mathcal{W}(-Q_{\perp})}{|Q_{\perp}|^4}\right) \,.
\label{eqn:second}
\eea
We have two regions of manifest IR divergence here: $k_{\perp}\rightarrow0$ and $ k_{\perp}\rightarrow - Q_{\perp}$. The first term in Eq.~(\ref{eqn:second}) is symmetric under the interchange ${\bs k}_{\perp} \leftrightarrow -{\bs Q}_{\perp}-{\bs k}_{\perp}$. So the two regions have the same singular behavior.
Expanding about these two singular regions (we only need to expand around one of them and then multiply by two), we obtain the leading singularity (LS):
\bea
 G^{(2)}(Q^-,Q_{\perp},t)\Big|_{\ma{LS}} &=& 2 \pi^2 t^2\left(  2\int_0^{|k_{\perp}|\ll |Q_{\perp}|} \diff^2 k_{\perp} \frac{\mathcal{W}(k_{\perp})}{|k_{\perp}|^4} \frac{\mathcal{W}( -{Q}_{\perp})}{|{Q}_{\perp}|^4}- 2 R(Q) \frac{\mathcal{W}(-Q_{\perp})}{|Q_{\perp}|^4}\right) \,.\nn\\
\eea
Given the form of $R(Q)$ in Eq.~(\ref{eqn:rate_form}), we see that the leading IR singularity cancels out. However, we can still have subleading IR singularities that do not cancel. To see this more explicitly, we can set $\mathcal{W}(k_{\perp})$ to be a constant since it has a very mild dependence on $k_{\perp}$. We can then write 
\bea 
 G^{(2)}(Q^-,Q_{\perp},t)&=&  2\pi^2 t^2 \bigg(  2\mathcal{W}^2\int \frac{|k_{\perp}|\diff|k_{\perp}|\diff \phi}{|k_{\perp}|^4} \frac{1}{(|Q_{\perp}|^2+|k_{\perp}|^2+2 |Q_{\perp}||k_{\perp}|\cos \phi)^2} \nn\\
 && - 2 R(Q)\frac{\mathcal{W}}{|Q_{\perp}|^4}
 \bigg)
\eea  
We can then expand in terms of small $|k_{\perp}|$. As before, we have exploited the symmetry of ${\bs k}_{\perp} \leftrightarrow -{\bs Q}_{\perp}-{\bs k}_{\perp}$ to account for both singular regions. After expanding, we can do the angular integral to obtain
\bea
 G^{(2)}(Q^-,Q_{\perp},t) &\approx& 4\pi^2 t^2 \mathcal{W}^2 \bigg( \int \frac{|k_{\perp}|\diff|k_{\perp}|}{|k_{\perp}|^4} \bigg[\frac{2\pi}{|Q_{\perp}|^4}- \frac{4\pi|k_{\perp}|^2}{|Q_{\perp}|^6}\bigg]-\int \frac{|k_{\perp}|\diff|k_{\perp}|}{|k_{\perp}|^4}  \frac{2\pi}{|Q_{\perp}|^4}\bigg) \,.\nn\\
\eea
As we can see, the leading singularity cancels out, but a subleading logarithmic singularity remains. The final measurement constrain $Q_\perp$ can only cut-off the leading IR singularity but not the subleading one. This is connected with the Markovian approximation, as will be explained in the following.

The $\ml{W}$ term is proportional to $\alpha_s^2$. So the singularity first appears at $\ml{O}(\alpha_s^4)$. One might think that perhaps we are missing some pieces and if we include terms at higher order in the coupling constant, when deriving our master equation, this subleading singularity would be cured. But we can immediately see a higher order term would only contribute to $\ml{O}(\alpha_s^6)$ at $\ml{O}(t^2)$ and hence cannot cancel out our IR divergence at $\ml{O}(\alpha_s^4)$. This singularity must therefore have a physical origin.

The key point here is that we are working in the Markovian approximation in which all coherence is lost between successive interactions with the medium. Therefore we can treat successive interactions as products of independent scatterings. The only constraint that we are imposing is that the $total$ transverse momentum accumulated by the collinear parton should be of the order of $Q_{\perp}$. The value of $Q_{\perp}$ is determined by the final measurement we are conducting, and is assumed to be $\sim T$. This implies that if the jet interacts more than once with the medium, it is possible for the jet to accumulate almost $Q_{\perp}$ transverse momentum in one interaction and very little in all the others. All the other interactions are therefore independent scatterings in which little or no transverse momentum is exchanged. For these scatterings, our power counting does not apply. We must instead use the EFT in the regime with $ \lambda \sim m_D/Q$, making it sensitive to the scale $m_D$. Therefore we have to put in a mass regulator such as the Debye mass $m_D $ to obtain a finite sensible result. The Debye mass defines the interaction range of these scatterings with little transverse momentum transfer.

This also suggests that in the Markovian approximation, we need smooth transition through all the EFTs from the scale $m_D$ up to the scale $Q_\perp\sim T$. The leading singularities discussed above cancel out at all orders in the expansion of Eq.~(\ref{eqn:G_expansion}). We sketch a proof for this in Appendix~\ref{sec:singularity}. If we work to higher orders in the expansion, we observe that some of the subleading singularities also cancel out. So we conjecture that at the $n$-th order, we are only left with a $\ln^{n-1}$ singularity. If this is correct, then our solution is resumming a logarithmic series.

\subsection{Numerical Results}
\label{sect:num}
We will calculate the density of states with $Q_\perp \neq 0$. More specifically, we will compute Eq.~(\ref{eqn:def_G}) numerically.
We will cut the IR divergence by introducing a gluon mass at finite temperature, which is the Debye screening mass 
\bea
\label{eqn:mD}
m_D^2 = \frac{1}{3}\Big(C_A+\frac{N_f}{2}\Big) g^2T^2 \,,
\eea
in which $C_A=N_c=3$ and $N_f=3$ is the number of active quark flavors in the QGP (we will assume the strange quark is massless for simplicity). In the following calculations, we will replace $ |k_\perp|^4$ in the denominator in Eq.~(\ref{eqn:sudakov}) with $(|k_\perp|^2+m_D^2)^2$.

The results shown in (\ref{eqn:def_G}) depend on time $t$, the temperature $T$ of the QGP and the transverse momentum $Q_\perp$ of interest. We will do the following scaling
\bea
\hat{t} &=& Tt\,, \ \ \ \ \, \hat{R} = \frac{R}{T}\,, \ \ \ \ \ \, \hat{\widetilde{K}} =\frac{\widetilde{K}}{T}\,, \nn\\
\hat{Q}_\perp &=& \frac{Q_\perp}{T}\,, \ \ \ \hat{r}_\perp = T r_\perp\,, \ \ \hat{k}_\perp =\frac{k_\perp}{T}\,, \nn\\
\hat{p}_\perp &=& \frac{\hat{p}_\perp}{T}\,, \ \ \ \ \hat{p}^- = \frac{p^-}{T}\,, \ \ \ \hat{m}_D = \frac{m_D}{T} \,,
\eea
so that the results at different temperatures fall onto a ``universal'' curve:
\bea
G(Q^-,Q_{\perp},t) &=& \frac{1}{T^2} \hat{G}(Q^-,\hat{Q}_{\perp},\hat{t}) \,.
\eea
The $\hat{G}$ function is given by
\bea
\hat{G}(Q^-,\hat{Q}_{\perp},\hat{t}) = 2\pi\int |\hat{r}_\perp| \diff|\hat{r}_\perp| J_0(|\hat{r}_\perp| |\hat{Q}_\perp| ) \Big( e^{-\hat{R}\hat{t} + \hat{\widetilde{K}}\hat{t}} - e^{-\hat{R}\hat{t}}\Big) \,,
\eea
in which the rate and kernel are given by
\bea
\label{eqn:hatR}
\hat{R} &=&\frac{4\alpha_s^2N_fC_FT_F}{\pi^2}\int \frac{|\hat{k}_\perp|\diff |\hat{k}_\perp|}{(|\hat{k}_\perp|^2+\hat{m}_D^2)^2} \hat{\ml{W}}(\hat{k}_\perp)  \\
\label{eqn:hatK}
\hat{\widetilde{K}} &=&\frac{4\alpha_s^2N_fC_FT_F}{\pi^2}\int \frac{|\hat{k}_\perp|\diff |\hat{k}_\perp|}{(|\hat{k}_\perp|^2+\hat{m}_D^2)^2}  J_0(|\hat{r}_\perp||\hat{k}_\perp|) \hat{\ml{W}}(\hat{k}_\perp) \\
\label{eqn:hatW}
\hat{\ml{W}}(\hat{k}_\perp) &=& \int  \diff |\hat{p}_\perp| \diff \hat{p}^- \diff \phi\, \frac{|\hat{p}_\perp|^3}{(\hat{p}^-)^2} \hat{n}_F\Big(\frac{(\hat{p}^-)^2+|\hat{p}_\perp|^2}{2\hat{p}^-}\Big)  \\
&\times&  \bigg[1-\hat{n}_F\Big(\frac{ (\hat{p}^-)^2 (|\hat{p}_\perp|^2 + |\hat{k}_\perp|^2 + 2|\hat{p}_\perp||\hat{k}_\perp|\cos\phi) +|\hat{p}_\perp|^4}{2|\hat{p}_\perp|^2\hat{p}^-}\Big)  \bigg] \,.
\eea
Here $\hat{n}_F(x) = (e^x+1)^{-1}$. The $\hat{G}$ function seems to be independent of the temperature $T$. But it still depends on $T$ through the coupling constant $\alpha_s$. The determination of $\alpha_s$ relies on a scale. If we scale everything by $T$, then $\alpha_s$ will depend on $T$.

\begin{figure}[h]
    \centering
    \begin{subfigure}[t]{0.5\textwidth}
        \centering
        \includegraphics[height=2.0in]{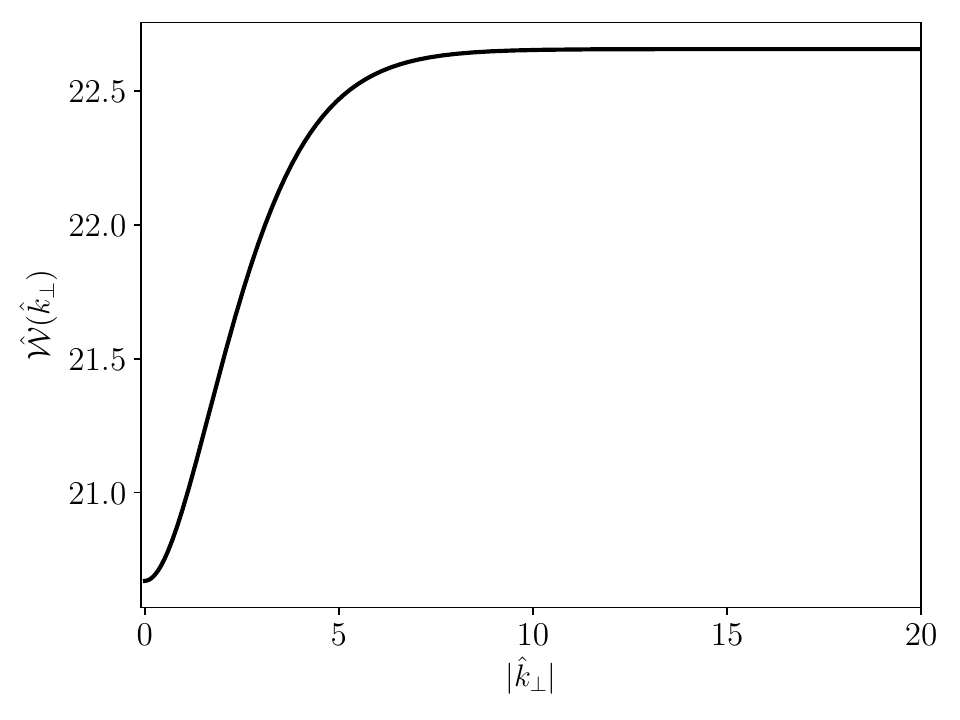}
        \caption{$\hat{\ml{W}}(\hat{k}_\perp)$.}
    \end{subfigure}%
    ~
    \begin{subfigure}[t]{0.5\textwidth}
        \centering
        \includegraphics[height=2.0in]{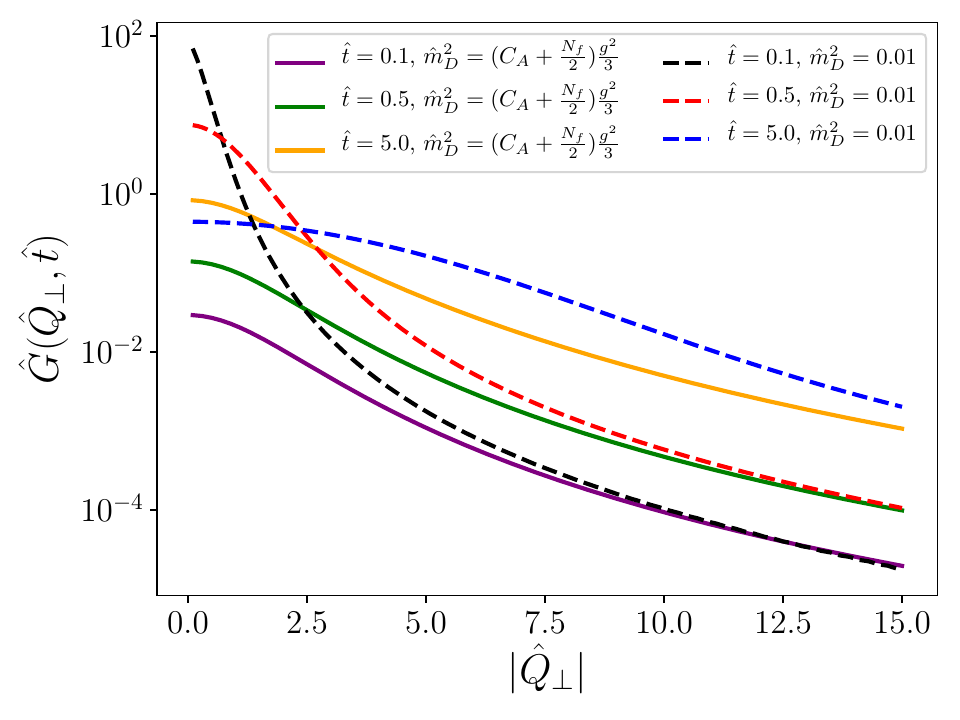}
        \caption{$\hat{G}(\hat{Q}_\perp,\hat{t})$.}
    \end{subfigure}
    \caption{Numerical results of $\hat{\ml{W}}$ and $\hat{G}$ with constant $\alpha_s=0.3$.}
    \label{fig:1}
\end{figure}

The numerical results of $\hat{\ml{W}}$ and $\hat{G}$ with constant $\alpha_s = 0.3$ are plotted in Fig.~\ref{fig:1}. Running coupling effect will be studied in the future. The function $\hat{\ml{W}}$ does not vary significantly as $|\hat{k}_\perp|$ changes. For $|\hat{k}_\perp|>10$, $\hat{\ml{W}}$ is almost a constant. In the plot of $\hat{G}$, we choose two different Debye screening masses to demonstrate the sensitivity to the IR scale: One is Eq.~(\ref{eqn:mD}) and for $\alpha_s=0.3$, $\hat{m}_D^2\approx5.65$; The other is $\hat{m}_D^2 = 0.01$. The curves of $\hat{G}$ are normalized according to Eq.~(\ref{eqn:final}):
\be
\label{eqn:Ghat_norm}
\int |\hat{Q}_\perp| \diff |\hat{Q}_\perp|  \hat{G}(|\hat{Q}_\perp|, \hat{t}) = 2\pi \big(1- e^{-\hat{R}\hat{t}} \big) \,.
\ee
The value of the Debye mass has a significant effect on the broadening rate of the jet. For the smaller Debye mass, the distribution is gradually broadened from the initial peak at the origin. For the larger Debye mass, the shape of $\hat{G}$ saturates fast and the only change is the normalization, which is given by Eq.~(\ref{eqn:Ghat_norm}). This is intuitively clear from the fact that for a value of $\hat{m}_D$ much greater than $|\hat{Q}_\perp|$ , the $|\hat{Q}_\perp|$ scale becomes irrelevant for the IR physics. Then the shape of the curve is fixed and its amplitude merely scales with time. We also notice that when $|\hat{Q}_\perp| \gg \hat{m}_D$, the results of $\hat{G}$ are less sensitive to the value of the Debye mass, where the Debye mass becomes irrelevant. Our simple calculations confirm the physical picture that the transverse momentum distribution of a collinear parton broadens as the parton traverses the QGP.

This is still only a partial picture since we have not included other effects such as vacuum parton shower evolution, medium-induced splitting and non-zero chemical potential. So a comparison with data will only be meaningful when these corrections are included in our evolution equation.

\subsection{Comparison with Previous Work}
The solution to the Markovian master equation (\ref{eqn:sol_to_master621}) agrees formally with Eq.~(2.15) of Ref.~\cite{DEramo:2012uzl}. To match the normalization used in Ref.~\cite{DEramo:2012uzl}, we need to choose $f(Q^-) = (2\pi)^2$. The exponent structure in the solution comes from resumming length enhanced diagrams in Ref.~\cite{DEramo:2012uzl} while in our case it shows up after solving the master equation. The solution resums multiple scattering, as also explained in Ref.~\cite{DEramo:2012uzl}. More specifically, the multiple scattering is incoherent. What Ref.~\cite{DEramo:2012uzl} calls $\ml{W}_{\ml{R}}^{(2)}$ corresponds to the Sudakov factor $S(Q,{\bs r}_\perp)= -R(Q) + \widetilde{K}(Q, -r_{\perp})$ we defined earlier. The diagrammatic interpretation of $\ml{W}_{\ml{R}}^{(2)}$ in Ref.~\cite{DEramo:2012uzl} exactly matches the structure of the Lindblad equation.

Next we compare the transverse momentum distribution after single scattering in our study and those calculated in Refs.~\cite{DEramo:2012uzl,DEramo:2018eoy}. This quantity is defined as $P_\ma{single}$ in Ref.~\cite{DEramo:2018eoy}. In our notation, this quantity is given by
\bea
P_\ma{single}(Q_\perp) = \int \diff^2r_\perp e^{-i{\bs r}_\perp \cdot {\bs Q}_\perp } \Big[ \widetilde{K}(Q, -r_\perp) - R(Q) \Big] t \,.
\eea
Our EFT has been designed with an expansion in the small parameter $Q_{\perp}/Q$ along with the condition $Q_{\perp} \gg m_D$. It is therefore valid for a wide range in the temperature $T$. It is possible to obtain simple analytic expressions in certain limiting hierarchies. To compare with earlier literature, we will focus on two kinematic regions: $|Q_\perp| \gg T$ and $T \gg |Q_\perp| \gg m_D$. Our EFT framework is still valid in these two limiting cases (the only difference is that the power counting parameter $\lambda$ is now given by $Q_\perp/Q^-$ rather than $T/Q^-$).
For $Q_\perp \neq 0$, neglecting the Debye mass, we find
\bea
\label{eqn:limits}
P_\ma{single}(Q_\perp) = (2\pi)^2 t \frac{\ml{W}(-Q_\perp)}{|Q_\perp|^4} = t \frac{8\alpha_s^2  N_f C_F T_F}{\pi}  \frac{T^3}{|Q_\perp|^4} \hat{\ml{W}}(-\hat{Q}_\perp) \,,
\eea
where $4\pi \alpha_s = g^2$ (note the difference in the definition of $\ml{W}$ in Eq.~(\ref{eqn:kernel_form}) and $\hat{\ml{W}}$ in Eq.~(\ref{eqn:hatK})).
Since here we only focus on the Glauber scattering between a collinear quark and a soft quark, we will only compare the results relevant for quarks. The results of Ref.~\cite{DEramo:2018eoy} for quark-quark scattering in these two regions can be written as
\bea
P_\ma{single}(Q_\perp) = \begin{cases}
    t g^4 C_F N_f \frac{3 \zeta(3)}{2\pi^2} \frac{T^3}{|Q_\perp|^4}
    ,&  |Q_\perp| \gg T \\[3pt]
    t g^4 C_F N_f \frac{1}{6}\frac{T^3}{|Q_\perp|^4},
    & T \gg |Q_\perp| \gg m_D
\end{cases} \,.
\eea
Our result (\ref{eqn:limits}) agrees with Ref.~\cite{DEramo:2018eoy} in these two limits. This can be easily checked using the numerical result of $\hat{\ml{W}}$ shown in Fig.~\ref{fig:1} or referring to the limiting formulas~(\ref{eqn:limit1}, \ref{eqn:limit2}) in the Appendix~\ref{sec:T_Correlator}. 

\section{Conclusions}
\label{sec:conclusions}

In this paper, we take the first step towards developing an effective field theory description for energetic jets traversing a region of the QGP medium. We treat the jet as an open quantum system interacting with a QGP environment in thermal equilibrium at constant temperature $T$. For now we restrict ourselves to the case when the scale $T$ is perturbative, which is valid at high temperature. We focus on the transverse momentum broadening in this paper and neglect parton splitting. The interaction between the jet and the medium is encoded in effective operators mediated by a Glauber momentum exchange, which captures the dominant forward scattering regime. Tracing out the degrees of freedom of the environment yields a Lindblad evolution equation for the density matrix of the jet. The Lindblad equation turns into a master equation in the Markovian limit if the initial jet density matrix is diagonal in the transverse momentum space and a final projective measurement onto a specific transverse momentum $Q_\perp$ is imposed.

Given that the Glauber mode connects the subsystem and the thermal QGP only via its transverse momentum, we observe that we can analytically solve the time evolution of the master equation by going to the impact parameter space. We work in the scale hierarchy $ Q_{\perp} \gg m_D \gg \Lambda_{QCD}$. The EFT works for a wide range of high temperatures and the expansion parameter for our EFT description is set by $\lambda \sim Q_{\perp}/Q \ll 1$. We demonstrate that the leading singularities cancel out but subleading singularites still exist, which we conjecture are powers of logarithms to all orders. We surmise that the residual IR singularity is a consequence of the Markovian approximation used to describe the density matrix evolution. We calculate numerically the transverse momentum distribution by cutting off the IR singularity with a gluon Debye mass. The distribution becomes broadened as time increases. Our results agree with those  previously derived in literature in various limiting regimes of our EFT.

Our ultimate goal is to understand the jet quenching inside a dynamically evolving QGP in a theoretically controlled way. Looking forward, there are several questions that we would like to address using our current approach. The most urgent one is to incorporate the effects of vacuum shower and other medium-induced effects (such as the medium-induced splitting) systematically. Furthermore, non-Markovian effect of the jet dynamics inside the QGP (such as the LPM effect\footnote{At least three time scales are involved in the LPM effect: the environment correlation time $\tau_E\sim1/T$, the Glauber exchange time scale $\tau_G\sim1/(\alpha_s T)$ and the formation time of the radiated gluon $\tau_F\sim \sqrt{\frac{x(1-x)E}{T}} \tau_G$, where $xE$ denotes the energy of the radiated gluon. The LPM effect is important when $\tau_F \gtrsim \tau_G$, when multiple Glauber exchanges can happen during the formation of the radiated gluon. These Glauber exchanges have to be resummed coherently in the amplitude level and their interference will lead to a suppression of the radiation. Thus, these Glauber exchanges cannot be treated as independent scatterings, and cannot be resummed in the Markovian master equation. One may construct effective operators in the Hamiltonian for the LPM modified radiation. If $\tau_F\gg \tau_E$ is valid, then each LPM modified gluon radiation can still be treated as independent processes, i.e., Markovian.}) should be investigated in our framework. We can also apply our formalism to study a heavy quark jet traveling through the medium. For the heavy quark jet, we need to first construct effective field theory for the forward scattering of a boosted heavy quark and then use those operators in our open quantum system formalism. Finally, we should explore how to define and compute jet substructure observables in our formalism and how to incorporate the wake of a jet into the formalism. Finally, we want to understand whether universal nonperturbative effects can be suitably parametrized in our formalism and then extracted from experiment. Understanding these questions will help us make better use of jets as probes of the QGP.

\acknowledgments
We thank Krishna Rajagopal, Iain Stewart and Yi Yin for inspiring discussions. This work is supported by the Office of Nuclear Physics of the U.S. Department of Energy under Contract DE-SC0011090 and Department of Physics, Massachusetts Institute of Technology.

\appendix
\section{Wightman Functions in Thermal Bath}
\label{sec:T_Correlator}
We will use the imaginary time formalism to compute the finite temperature correlation function of the soft operators. The operator for the soft quark current is given by (\ref{eqn:softO}):
\bea
\mathcal{O}_s^{q_n A}= 8\pi\alpha_s\frac{1}{\mathcal{P}_{\perp}^2}\bar\psi^n_s  T^A \frac{\slashed{n}}{2}\psi^n_s \,,
\eea 
where the soft quark operator $\psi_s^n$ is dressed with a soft Wilson line $\psi_s^n = S_n^\dagger \psi_s$. In our weak coupling calculation, the soft Wilson line can be dropped at leading order.
In the imaginary time formalism, we first compute the correlator in the Euclidean space and then analytically continue to the Minkowski space. The Euclidean correlator in momentum space is defined by
\bea
D_E^{AB}(K) = \int_0^{\beta}\diff\tau \int \diff^3 {\bs x} \, e^{i K \cdot X} \langle  \mathcal{O}_s^{q_n A}(X) \mathcal{O}_s^{q_n B}(0)\rangle\,,
\eea
where $\beta = 1/T$. The Euclidean coordinate is $X=(\tau=it, {\bs x})$, the Euclidean momentum is $K=(k_\ell, {\bs k})$ and $K\cdot X = k_\ell\tau -{\bs k}\cdot {\bs x}$. The Matsubara frequency for bosonic operator (our soft quark current operator is bosonic even though it comprises quark fields) is $k_\ell=2\ell\pi T$ where $\ell$ is an integer. Plugging the soft quark current into the correlator leads to
\bea
 D_E^{AB}(K) &=& -(8\pi\alpha_s)^2 \int_0^{\beta}\diff\tau \int \diff^3 {\bs x} \, T\sum_n T\sum_m \int \frac{\diff^3 {\bs p}}{(2\pi)^3} \int \frac{\diff^3 {\bs q}}{(2\pi)^3}e^{i (K+P-Q)\cdot X} \nn \\ 
 &\times&\frac{1}{[({\bs p}_{\perp}-{\bs q}_{\perp})^2]^2} \Tr\Big[\frac{-i \slashed P}{P^2}\frac{\slashed{n}}{2}\frac{-i \slashed Q}{Q^2}\frac{\slashed{n}}{2}\Big]\Tr[T^{A}T^{B}] \,,
\eea
where $P = (p_n, {\bs p})$ and $ Q=(q_m, {\bs q})$. Here $p_n=(2n+1)\pi T$ and $q_m=(2m+1)\pi T$ are the Matsubara frequencies for fermionic operators. The notations here are Euclidean: $P^2 = p_n^2+{\bs p}^2$, $\slashed{P} = \gamma^0_E p_n + \gamma^i_E p_i = \gamma^0p_n -i\gamma^ip_i = \gamma^0p_n +i\gamma^ip^i$.
The overall minus sign comes from the fermion loop. After the integral over the Euclidean coordinates, we obtain
\bea
 D_E^{AB}(K) = -(8\pi\alpha_s)^2 T\sum_n \int \frac{\diff^3{\bs p}}{(2\pi)^3}  \frac{1}{[({\bs k}_{\perp})^2]^2} \Tr \Big[\frac{-i \slashed P}{P^2}\frac{\slashed{n}}{2}\frac{-i (\slashed K +\slashed P)}{(P+K)^2}\frac{\slashed{n}}{2}\Big] \Tr [T^{A}T^{B}] \,.
\eea
We now apply the trick known as the Saclay method to sum over the Matsubara frequencies. To that end, we introduce a Kronecker delta function by writing 
\bea
 D_E^{AB}(K) =-(8\pi\alpha_s)^2 \frac{\Tr [T^{A}T^{B}]}{[({\bs k}_{\perp})^2]^2} \int \frac{\diff^3 {\bs p}}{(2\pi)^3}  T\sum_{n} T\sum_m \beta \delta_{q_m,p_n+k_\ell} \frac{f(ip_n,ik_\ell, {\bs p},{\bs k})}{[p_n^2+E_1^2][q_m^2+E_2^2]}\,,\ \ \ \ 
\eea 
where the function $f$ contains all the trace factors in the numerator, with $ E_1 = |{\bs p}|$, $E_2= |{\bs p}+{\bs k}|$. Moreover, $p_n$ and $q_m$ are fermionic Matsubara frequencies. Then we can write
\bea 
 D_E^{AB}(K) &= & -(8\pi\alpha_s)^2\frac{\Tr[T^AT^B]}{[({\bs k}_{\perp})^2]^2}\int_0^{\beta} \diff\tau e^{-i \tau k_\ell}\int \frac{\diff^3{\bs p}}{(2\pi)^3} \nn\\
 &\times& \Bigg[T\sum_{n}e^{-i \tau p_n} \frac{f(ip_n,ik_\ell, {\bs p},{\bs k})}{p_n^2+E_1^2}\Bigg]\Bigg[T\sum_{m} \frac{e^{i \tau q_m}}{q_m^2+E_2^2}\Bigg] \,.
\eea
We can further simplify the above expression by using the following relations:
\bea
T\sum_{n}\frac{e^{- ip_n \tau}}{p_n^2+E_1^2} = \frac{n_F(E_1)}{2E_1}\Big[e^{(\beta-\tau)E_1}-e^{\tau E_1}\Big] \nn\\
T\sum_{n}\frac{ip_ne^{- ip_n \tau}}{p_n^2+E_1^2}=  \frac{n_F(E_1)}{2E_1}\Big[E_1e^{(\beta-\tau)E_1}+E_1e^{\tau E_1}\Big] \nn\\
T\sum_{n}\frac{(ip_n)^2e^{- ip_n \tau}}{p_n^2+E_1^2}=  \frac{n_F(E_1)}{2E_1}\Big[E_1^2e^{(\beta-\tau)E_1}-E_1^2e^{\tau E_1}\Big] \,,
\eea
where 
$n_F(E)=(e^{\beta E}+1)^{-1}$ is the Fermi-Dirac distribution. After applying the summation formulas, the integral over $\tau$ can be easily done.

Once we obtain the correlator in the imaginary time formalism, we can obtain all the real time Green's functions and the spectral function via analytic continuation. First, the spectral function can be obtained by 
\bea
\label{eqn:spectral}
\rho^{AB}(k^0,{\bs k}) &=& \frac{1}{i} Disc \, D^{AB}_E(k_n \rightarrow -ik^0, {\bs k}) \nn\\
&=& -i \Big( D^{AB}_E(-i[k^0+i0^+], {\bs k} )-D^{AB}_E(-i[k^0-i0^+], {\bs k})\Big) \,.
\eea
In our case, Eq.~(\ref{eqn:spectral}) leads to
\bea
\rho^{AB}(k) &=&   (8\pi\alpha_s)^2\frac{T_F\delta^{AB}}{[({\bs k}_{\perp})^2]^2}\int \frac{\diff^3{\bs p} }{(2\pi)^34E_1 E_2} \nn\\
&\times& \bigg[ 2 n\cdot p n \cdot (p+ k) \Big(1-n_F(E_1) -n_F(E_2)\Big) 2\pi\delta(k^0+E_1+E_2)  \nn\\
&+& 2 n\cdot p n \cdot (p+ k)\Big( n_F(E_1)-n_F(E_2)\Big) 2\pi\delta(k^0+E_1-E_2) \nn\\
&+& 2 \bar n \cdot p(\bar n\cdot p - n\cdot k) 
\Big( n_F(E_2)-n_F(E_1)\Big) 2\pi\delta(k^0+E_2-E_1) \nn\\
&-& 2 \bar n \cdot p(\bar n\cdot p - n\cdot k)\Big(1-n_F(E_2)-n_F(E_1)\Big) 2\pi\delta(k^0-E_1-E_2)\bigg]
\eea
The four terms here present four different scattering processes respectively. The ones of our interest are the second and third term, which correspond to scattering processes with one incoming and one outgoing soft quark. The other two terms correspond to processes with either two incoming soft quarks or two outgoing soft quarks. In fact, these two terms (the term proportional to $\delta(k^0+E_1+E_2)$ or $\delta(k^0-E_1-E_2)$) do not contribute at the order we are working here, since the gluon exchanged in these processes scales as a soft mode rather than a Glauber mode. In our power counting, the soft mode scales as $p_s\sim Q(\lambda,\lambda,\lambda)$ and the sum of any two soft modes scales similarly. If the exchanged gluon is soft, the collinear particle will become off-shell. So this process is suppressed and we can drop them. At the same time, we can use our power counting to drop the term $n\cdot k$ (which is Glauber and scales as $\sim \lambda^2$) when compared with $\bar{n}\cdot p$ or $n\cdot p$ (which are soft and scale as $\sim \lambda$)
\bea
\rho^{AB}(k) &=&   (8\pi\alpha_s)^2\frac{T_F\delta^{AB}}{[({\bs k}_{\perp})^2]^2}\int \frac{\diff^3{\bs p} }{(2\pi)^34E_1 E_2} \nn\\
&\times& \bigg[ 2 (n\cdot p)^2 \Big(n_F(E_1)-n_F(E_2)\Big) 2\pi\delta(k^0+E_1-E_2) \nn\\
&+& 2 (\bar n \cdot p)^2 \Big( n_F(E_2)-n_F(E_1)\Big)2\pi\delta(k^0+E_2-E_1)\bigg] \,.
\eea

If we define $\rho^{AB}(k) = \rho(k) \delta^{AB}$, we can write the Wightman function $D_>^{AB}(k) = D_>(k)\delta^{AB}$ as
\bea
D_>(k) = \big( 1+n_B(k^0) \big) \rho(k) \,. 
\eea
Plugging the spectral function gives
\bea
D_>(k)&=& (8\pi\alpha_s)^2\frac{T_F}{[({\bs k}_{\perp})^2]^2}\int \frac{\diff^3 {\bs p}}{(2\pi)^3 4E_1 E_2} \nn\\
&\times& \Big[ 2 (n\cdot p)^2 n_F(E_1)\big(1-n_F(E_2)\big) 2\pi \delta(k^0+E_1-E_2) \nn\\
&+& 2 (\bar n \cdot p)^2 n_F(E_2)\big(1-n_F(E_1)\big)2\pi \delta(k^0+E_2-E_1)\Big] \,,
\eea
where $n_F$ appears for the initial state while $1-n_F$ shows up in the final state. This is the standard Pauli blocking in quantum statistics.
In order to apply the power counting in $k$, we need to express everything in light-cone coordinates. To that end, we make a change of variables ${\bs p} \rightarrow - {\bs p} $ in the second term and introduce a dummy four-momentum variable $q$ to write
\bea
D_>(k)&=& (8\pi\alpha_s)^2\frac{4\pi T_F}{[({\bs k}_{\perp})^2]^2}\int \frac{\diff^4p \diff^4q}{(2\pi)^3} \delta^+(p^2)\delta^+(q^2) 2 (n \cdot p)^2n_F(p^0) \big(1-n_F(q^0) \big)\delta^4(k+p-q) \,.\nn\\
\eea
The integral over $q$ gives
\bea
D_>(k)&=& (8\pi\alpha_s)^2\frac{4\pi T_F}{[({\bs k}_{\perp})^2]^2}\int \frac{\diff^4p}{(2\pi)^3} \delta^+(p^2)\delta^+\big((k+p)^2\big) 2 (n \cdot p)^2n_F(p^0) \big(1-n_F(k^0+p^0) \big) \nn\\
&\equiv& (8\pi\alpha_s)^2\frac{4\pi T_F}{(2\pi)^3[({\bs k}_{\perp})^2]^2} I(k^-, k_\perp) \,,
\eea
where we define the integral as $I$. Now our task is to simplify the integral 
\bea
 I(k^-, k_\perp) &\equiv& \int \diff^4p \, \delta^+(p^2)\delta^+\big((k+p)^2\big) 2 (n \cdot p)^2n_F(p^0) \big(1-n_F(k^0+p^0) \big)  \nn\\
&=& \int |p_\perp| \diff |p_\perp|\diff p^- \diff p^+ \diff \phi \delta(p^-p^+-|p_\perp|^2) \delta\big( (p^-+k^-)p^+ - |p_\perp+k_\perp|^2\big) (p^+)^2 \nn \\
\label{eqn:appendix_I}
&\times& \Theta(p^-+p^+) \Theta(p^-+k^-+p^+)
 n_F\Big(\frac{p^-+p^+}{2}\Big) \bigg[1-n_F\Big(\frac{p^-+k^-+p^+}{2}\Big)  \bigg] \,,\ \ \ \ \ \ \ 
\eea
where we have dropped $k^+$ according to our power counting.

For later convenience, we now calculate 
\bea
&&\int \frac{\diff k^-}{2\pi} I(k^-, k_\perp)  \nn\\
&=&\frac{1}{2\pi}\int |p_\perp| \diff |p_\perp| \diff p^- \diff \phi\, \frac{(p^+)^2}{p^- p^+}  n_F\Big(\frac{p^-+p^+}{2}\Big) \bigg[1-n_F\Big(\frac{p^-+k^-+p^+}{2}\Big)  \bigg]\,,
\eea
where $p^->0$ and the values of $p^+$ and $k^-$ are fixed by integrating over the two delta functions:
\bea
p^+ &=& \frac{|p_\perp|^2}{p^-} \\
k^- &=& \frac{|p_\perp + k_\perp|^2}{p^+} - p^- =  p^- \frac{|p_\perp|^2 + |k_\perp|^2 + 2|p_\perp||k_\perp|\cos\phi }{|p_\perp|^2} - p^- \,.
\eea
In the limit $|k_\perp|\to 0$, we find $k^- \to0$, so
\bea
&&\lim_{|k_\perp|\to0}\int \frac{\diff k^-}{2\pi} I(k^-, k_\perp) \nn\\
\label{eqn:limit1}
&=& \int \diff |p_\perp|\diff p^-  \frac{|p_\perp|^3 }{(p^-)^2} n_F\Big(\frac{(p^-)^2+|p_\perp|^2}{2p^-}\Big) \bigg[1-n_F\Big(\frac{(p^-)^2+|p_\perp|^2}{2p^-}\Big) \bigg] = \frac{\pi^2}{3} \,.
\eea
In the limit $|k_\perp|\to \infty$, we find $k^- \to \infty$, so we can approximate the $(1-n_F)$ term by one to obtain
\bea
\label{eqn:limit2}
\lim_{|k_\perp|\to\infty}\int \frac{\diff k^-}{2\pi} I(k^-, k_\perp) =\int \diff |p_\perp|\diff p^-  \frac{|p_\perp|^3 }{(p^-)^2} n_F\Big(\frac{(p^-)^2+|p_\perp|^2}{2p^-}\Big) = 3\zeta(3) \,.
\eea

So far, we only considered one flavor of massless soft quark. If we assume the medium consists of $N_f=3$ massless soft quarks (we neglect the strange quark mass), we find after putting everything together
\bea
\label{eqn:D>k}
D_>(k^-,k_\perp)&=& N_f\frac{(8\pi\alpha_s)^2}{(2\pi)^3}\frac{4\pi T_F}{|k_{\perp}|^4}\times I(k^-,k_\perp) \,.
\eea


\section{Rate and Kernel}
In this appendix, we will explain the computation of the rate $R$ and the kernel $K$ for a collinear quark scattering with soft quarks of the medium.

\label{sec:rate_kernel}
\subsection{Rate}
From Eq.~(\ref{eqn:rate}), we have the expression for the dissipation rate that appear in the final master equation
\bea
R(Q) = \int\frac{\widetilde{\diff q}}{2E_Q} \int \frac{\diff^4k}{(2\pi)^4} D_{>}^{AB}(k) \frac{1}{N_c}\bigg[\bar u_{n,s}(Q)\frac{\slashed{\bar n}}{2}\bar n \cdot q T^{A} T^B u_{n,r}(Q)\bigg](2\pi)^4\delta^4(Q-k-q)\,,\ \ \ \ \ 
\eea
where we restore the spin indexes $s,r$ here and we will show they are equal.
To simplify this, we use the fact that $ D_{>}^{AB} \propto \delta^{AB}$ and write 
\bea
 D_{>}^{AB}(k)  = \delta^{AB} D_{>}(k) \,.
\eea
Then we can work out the color factors
\bea
 D_{>}^{AB}(k) \bigg[\bar u_{n,s}(Q)\frac{\slashed{\bar n}}{2} \bar n \cdot q T^A T^B u_{n,r}(Q)\bigg]=D_{>}(k) C_F \bar n \cdot Q \bar n \cdot q \delta_{sr} \,,
\eea
where we average over the color of the incoming state with momentum $Q$. To obtain the result, we have used 
\bea
T^A T^A &=& C_F \mathbb{1}_{N_c\times N_c} \\
\bar u_{n,s}(Q)\frac{\slashed{\bar n}}{2}u_{n,r}(Q) &=& \delta_{sr}\bar n \cdot Q \,.
\eea
From now on, we will omit the spin indexes $s,r$. We are left with 
\bea
 R(Q) &=& C_F \int \widetilde{\diff q} \int \diff^4k D_{>}(k)\delta^4(Q-k-q)\bar n \cdot q  \nn\\
&=&  C_F \int \frac{\diff^4k}{(2\pi)^3} D_{>}(k) \delta^+ \Big( Q^-(Q^+-k^+)-({\bs Q}_{\perp}-{\bs k}_{\perp})^2 \Big) Q^- \,,
\eea
which we have used our power counting to expand away any power corrections. Here $Q$ is collinear and $k$ is Glauber. As we have shown in Appendix~\ref{sec:T_Correlator}, the $k^+$ dependence in $D_{>}(k)$ is dropped out since it is subleading in our power counting. Hence, we can easily do the integral over $k^+$ which leads to
\bea
  R(Q)&=& \frac{C_F}{2}  \int \frac{\diff^2 k_{\perp} \diff k^-}{(2\pi)^3}D_>(k^-,k_{\perp}) \,.
\eea
Plugging Eq.~(\ref{eqn:D>k}) into $R(Q)$ leads to
\bea
R(Q) &=& \frac{2\alpha_s^2N_fC_FT_F}{\pi^3} \int \frac{|k_\perp|\diff |k_\perp|}{|k_\perp|^4} \diff\phi_k \int  \diff |p_\perp| \diff p^- \diff \phi\, \frac{|p_\perp|^3}{(p^-)^2}  n_F\Big(\frac{(p^-)^2+|p_\perp|^2}{2p^-}\Big) \nn \\
&\times& \bigg[1-n_F\Big(\frac{ (p^-)^2 (|p_\perp|^2 + |k_\perp|^2 + 2|p_\perp||k_\perp|\cos\phi) +|p_\perp|^4}{2|p_\perp|^2p^-}\Big)  \bigg] \,,
\eea
where the integrand is independent of $\phi_k$ and the integral over $\phi_k$ can be done trivially.

\subsection{Kernel}
For the solution to the master equation, we also need to compute the kernel, which is defined by Eq.~(\ref{eqn:kernel_momentum}) in the transverse plane. Its expression can be obtained from Eqs.~(\ref{eqn:master0}, \ref{eqn:kernel} and \ref{eqn:kernel_momentum})
\bea
\widetilde{K}(Q,-r_\perp) &=& \int\frac{\widetilde{\diff q}}{2E_q}  \int \frac{\diff^4k}{(2\pi)^4} e^{-i{\bs k}_\perp\cdot {\bs r}_\perp} D_{>}^{AB}(k) \nn\\
&\times& \frac{1}{N_c}\bigg[\bar u_n(q)\frac{\slashed{\bar n}}{2}T^A u_n(Q)\bar u_n(Q)\frac{\slashed{\bar n}}{2}T^B u(q)\bigg](2\pi)^4\delta^4(Q+k-q) \,.
\eea
The evaluation of the kernel is almost the same as the rate, except for the extra phase $e^{-i {\bs k}_\perp \cdot {\bs r}_\perp}$. The result can be written as
\bea
\widetilde{K}(Q,-r_\perp) &=& \frac{C_F}{2}  \int \frac{\diff^2 k_{\perp} \diff k^-}{(2\pi)^3} e^{-i{\bs k}_\perp\cdot {\bs r}_\perp} D_>(k^-,k_{\perp}) \\
&=& \frac{2\alpha_s^2N_fC_FT_F}{\pi^3} \int \frac{|k_\perp|\diff |k_\perp|}{|k_\perp|^4}
\diff\phi_k e^{-i| k_\perp| | r_\perp| \cos\phi_k}
\int  \diff |p_\perp| \diff p^- \diff \phi\, \frac{|p_\perp|^3}{(p^-)^2}   \nn \\
&\times& n_F\Big(\frac{(p^-)^2+|p_\perp|^2}{2p^-}\Big) \bigg[1-n_F\Big(\frac{ (p^-)^2 (|p_\perp|^2 + |k_\perp|^2 + 2|p_\perp||k_\perp|\cos\phi) +|p_\perp|^4}{2|p_\perp|^2p^-}\Big)  \bigg] \,, \nn
\eea
where $\phi_k$ is the relative angle between ${\bs k}_\perp$ and ${\bs r}_\perp$.


\section{Proof for Cancellation of Leading IR Singularities}
\label{sec:singularity}
Here we sketch a proof of the statement that the leading singularities cancel out at all orders in the expansion of $t$, which is the expansion in Eq.~(\ref{eqn:G_expansion}). We will use the method of mathematical induction. 
The term at $n+1$-th order in the expansion can be written as (we will drop the $Q$ dependence in relevant functions)
\bea
 G^{(n+1)}&=&\frac{t^{n+1}}{(n+1)!} \int \diff^2r_{\perp} e^{-i {\bs r}_{\perp}\cdot {\bs Q}_{\perp}}\left(\Big[\widetilde{K}(Q, -r_{\perp})  -  R(Q) \Big]^{n+1}-\Big[-R(Q)\Big]^{n+1}\right)\nn\\
&=& \frac{t^{n+1}}{(n+1)!} \int \diff^2r_{\perp} e^{-i {\bs r}_{\perp}\cdot {\bs Q}_{\perp}}\left(\Big[\widetilde{K}( -r_{\perp})  -  R \Big] \Big[\widetilde{K}( -r_{\perp})  -  R \Big]^{n} -\Big[-R\Big]^{n+1}\right)\nn\\
&=& \frac{t^{n+1}}{(n+1)!} \int \diff^2r_{\perp} e^{-i {\bs r}_{\perp}\cdot {\bs Q}_{\perp}} \bigg( \widetilde{K}( -r_{\perp}) \Big[\widetilde{K}( -r_{\perp})  -  R \Big]^{n}  \nn\\
&-& R \Big( \big[\widetilde{K}( -r_{\perp})  -  R \big]^{n} - \big[-R\big]^{n} \Big) \bigg)\nn\\
&=& \frac{t^{n+1}}{(n+1)!}\int \diff^2r_{\perp} e^{-i {\bs r}_{\perp}\cdot {\bs Q}_{\perp}}  \bigg( \widetilde{K}( -r_{\perp}) \Big( \big[\widetilde{K}( -r_{\perp})  -  R \big]^{n} - \big[-R\big]^{n} \Big) 
+ \widetilde{K}( -r_{\perp}) \big[-R\big]^n\nn\\
&-&R \Big( [\widetilde{K}( -r_{\perp})  -  R \big]^{n} - \big[-R\big]^{n} \Big) \bigg) \,.
\label{singular}
\eea 
We have now written the result in terms of the $n$-th order term. We now only need to consider new singularities that arise as we go from the $n$-th to $(n+1)$-th order term. We first check the first term in the expression above:
\bea
 &&\int \diff^2r_{\perp} e^{-i {\bs r}_{\perp}\cdot {\bs Q}_{\perp}}  \bigg( \widetilde{K}( -r_{\perp}) \Big( \big[\widetilde{K}( -r_{\perp})  -  R \big]^{n} - \big[-R\big]^{n} \Big) \bigg)\nn\\
&=& \int \diff^2r_{\perp} e^{-i {\bs r}_{\perp}\cdot {\bs Q}_{\perp}}  \bigg(\int \frac{\diff^2 k_{\perp} e^{-i {\bs k}_{\perp} \cdot {\bs r}_{\perp}} \ml{W}(k_\perp)}{|k_{\perp}|^4} \Big( \big[\widetilde{K}( -r_{\perp})  -  R \big]^{n} - \big[-R\big]^{n} \Big) \bigg) \,.
\label{singular1}
\eea
Eq.~(\ref{singular1}) may have two new singularities at leading order: The first is when $k_{\perp} \rightarrow 0$ and the second is when $k_{\perp} \rightarrow -Q_{\perp}$ (see the explicit second order calculation in the main text). In the first case the exponential is approximately unity and the expression reduces to 
\bea
  R \int \diff^2r_{\perp} e^{-i {\bs r}_{\perp}\cdot {\bs Q}_{\perp}} \Big( \big[\widetilde{K}( -r_{\perp})  -  R \big]^{n} - \big[-R\big]^{n} \Big) \,,
\eea
which cancels out with the third term in the last line of Eq.~(\ref{singular}). No other new leading order singularities can show up in the $k_\perp\to0$ region because all leading order singularities cancel out in the $n$-th term
\bea
\int \diff^2r_{\perp}e^{-i {\bs r}_{\perp}\cdot {\bs Q}_{\perp}} \Big( \big[\widetilde{K}( -r_{\perp})  -  R \big]^{n} - \big[-R\big]^{n} \Big) \,. 
\eea

In the second case when $k_{\perp} \rightarrow -Q_{\perp}$, the term
\bea
\int \diff^2r_{\perp} e^{-i {\bs r}_{\perp}\cdot {\bs Q}_{\perp}} \int \frac{\diff^2 k_{\perp} e^{-i {\bs k}_{\perp} \cdot {\bs r}_{\perp}} \ml{W}(k_\perp)}{|k_{\perp}|^4}  \big[\widetilde{K}( -r_{\perp})  -  R \big]^{n}
\eea
in Eq.~(\ref{singular1}) has no leading singularity since $\widetilde{K}( -r_{\perp})  -  R$ has only logarithmic singularity, which is subleading. So for the consideration of leading IR singularity, Eq.~(\ref{singular1}) in the second region $k_{\perp} \rightarrow -Q_{\perp}$ reduces to 
\bea
- \int \diff^2r_{\perp}e^{-i {\bs r}_{\perp}\cdot {\bs Q}_{\perp}} \widetilde{K}( -r_{\perp})  \big[-R\big]^{n}
\eea
which cancels out with the second term in the last line of Eq.~(\ref{singular}).


\end{document}